\documentstyle[eqsecnum,aps,preprint]{revtex}
\tightenlines

\begin{document}
\draft
\title{Vortex configurations in a Pb/Cu microdot with a \\
2$\times$2 antidot cluster}
\author{T. Puig\cite{byline}, E. Rosseel, L. Van Look, M.J. Van Bael, V.V.
Moshchalkov and Y. Bruynseraede}
\address{Laboratorium voor Vaste-Stoffysica en Magnetisme,\\
Katholieke Universiteit Leuven\\
Celestijnenlaan 200D, B-3001 Leuven, Belgium}
\author{R. Jonckheere}
\address{Inter-university Micro-Electronics Center\\
Kapeldreef 75, B-3001 Leuven, Belgium}
\date{\today}
\maketitle

\begin{abstract}
We present a detailed study of the transport properties of a superconducting
Pb/Cu microdot with a $2\times 2$ antidot cluster. The
superconducting-normal (S/N) phase boundary, critical currents and
current-voltage characteristics of this structure have been measured. The
S/N phase boundary as a function of field $B$ ($T_c(B)$) reveals an
oscillatory structure caused by the limited number of possible vortex
configurations which can be realized in these small clusters of pinning
centres (antidots). We have analyzed the stability of these configurations
and discussed the possible dissipation mechanisms using the critical current
($J_c(B)$) and voltage-current ($V(I)$) characteristics data. A comparison
of the experimental data of $T_c(B)$ and $J_c(B)$ with calculations in the
London limit of the Ginzburg-Landau theory confirms that vortices can indeed
be pinned by the antidots forming a cluster and that the ground-state
configurations of the vortices are noticeably modified by sending current
through the structure. The possibility of generating phase-slips as well as
motion of the vortices in the $2\times 2$ antidot cluster has also been
discussed.
\end{abstract}

\pacs{Pacs: 74.25.Dw, 73.23-b, 74.76-w}

\preprint{}

\narrowtext

\section{Introduction}

\label{sec:intro}

Laterally modulated superconductors have been intensively studied during the
past few years (for reviews see \cite{Mooij88,Pannetier91}). The interest
towards these materials was stimulated by the advances in the
microfabrication techniques which enabled the production of sub-micron
structures with the relevant dimensions of the order of the superconducting
coherence length, $\xi (T)$ and penetration depth $\lambda (T)$. These
microstructures impose certain constraints on the behaviour of the
superconducting order parameter by confining the superconducting condensate
within the sample's boundaries, including those introduced through lateral
nanostructuring. Superconducting wire networks and Josephson junction arrays
(JJA's) are some examples of the laterally modulated films, where the
applied magnetic field provides a continuously changing length scale which
can explore commensurability between the lattice of vortices and the
underlying artificially introduced lateral lattice \cite
{Giroud92,Vanderzant90,Vanderzant94}.

In networks, fluxoid quantization governs the properties and defines the
allowed quantum states. Circular currents (vortices) are induced around the
cells forming the structure, thus defining sets of specific vortex
configurations.

Recently, a new class of superconducting systems was introduced, consisting
of a regular pinning array formed by a lattice of antidots (i.e. submicron
holes) made in a type-II superconducting film \cite{Baert95,Bezryadin95}. At
temperatures close to $T_c$ superconductors with an antidot lattice behave
like weakly coupled wire networks \cite{Rosseel96a}, where vortex depinning
dominates the dissipation \cite{Giroud92,Vanderzant90,Vanderzant94}. Well
below the transition temperature, where the fluxoid quantization condition
implies flux quantization at the antidots, new features appear in the
magnetization \cite{Baert95,Moshchalkov96}, critical currents \cite
{Rosseel96b} and S/N phase boundaries \cite{Bezryadin95} which have been
attributed to the formation of multiquanta vortices at the antidots.

Taking into account the vortex-vortex interactions in samples with $\sim
10^6 $ antidots is, however, not a simple exercise due to the very large
number of the interacting vortices. From this point of view, $%
a\;microdot\;with\;an\;antidot\ cluster$ ($2\times 2$, $3\times 3$, etc.)
with a small number of interacting vortices is a very promising
''intermediate'' system between a single superconducting loop with a finite
strip width and a superconducting film with a huge array of antidots. At the
same time, using finite microdots with four antidots (the antidot cluster),
we still keep the most essential features of the vortex-vortex interactions
in presence of antidots. The reduced number of interacting vortices
simplifies exact calculations which can also be extrapolated for the
analysis of the vortex behaviour in substantially larger antidot arrays.

In this paper, we study the transport properties of such a model
superconducting microsquare containing four antidots. Further on, for
simplicity, we shall call this structure a ''$2\times 2$ antidot cluster''.
This system can be considered as four unit cells of the regular square
lattice of antidots and the initial simplicity makes the $2\times 2$ antidot
cluster a good candidate for achieving an insight into the stable vortex
configurations expected for larger systems. Only a very limited number of
vortex configurations are allowed for the $2\times 2$ antidot cluster and
therefore both the experimental results are less difficult to be interpreted
and also a computational analysis is easier to perform. Additionally, the
limited number of vortex configurations makes these structures also
interesting for flux logic applications \cite{Puig96}.

In this paper we focus on measurements of the superconducting/normal (S/N)
phase boundary, the magnetoresistance, the critical currents and the $V(I)$
characteristics of a superconducting microsquare with four antidots. These
experimental results have been compared with calculations in the London
limit of the Ginzburg-Landau theory and in the framework of the de
Gennes-Alexander model\cite{DeGennesalexander}. The comparison has
demonstrated that several well defined vortex configurations can be induced
in these samples by properly tuning the magnetic field and the temperature.
At very low transport currents, only four of these configurations are
stable. By increasing the transport current, however, other flux phases,
which are unstable without transport current, can be realized. Based on the $%
V(I)$ characteristics data, we also consider how the occupation of antidots
by vortices influences the onset of dissipation.

The paper is organized as follows. Section~\ref{sec:exp} describes the
sample preparation, the experimental techniques and characteristic
superconducting parameters of the Pb/Cu 2$\times $2 antidot cluster. In
section~\ref{sec:phasbo}, we present the results on the S/N phase boundary $%
T_c(B)$ for the $2\times 2$ antidot cluster and a reference superconducting
microsquare without antidots. The calculation of the S/N phase boundary for
the 2$\times $2 antidot cluster in the London limit and in the framework of
the de Gennes-Alexander model \cite{DeGennesalexander} is described and
compared with the experimental data. The discussion of possible effects of
disorder is also presented. Section~\ref{sec:critcur} comprises the
experimental results of the critical currents and $V(I)$ characteristics.
The $V(I)$ data are used to determine the dissipation related to the vortex
motion in the 2$\times $2 antidot cluster.


\section{Experimental}

\label{sec:exp}

\subsection{Sample preparation and characterization}

\label{sec:prepchar}

The 2$\times $2 antidot clusters consisting of microsquares of 2$\times $2 $%
\mu m^2$ with four antidots (i.e. submicron holes of 0.53$\times $0.53 $\mu
m^2$ and center to center distance of 1 $\mu m$) were written by e-beam
lithography in positive PMMA resist onto SiO$_2$ substrates. Afterwards, a
bilayer consisting of 50 nm thick Pb and 17 nm thick Cu was evaporated in
one single run in a MBE apparatus at pressures of 5$\times 10^{-8}$ Torr.
The SiO$_2$ substrates were N$_2$ cooled during evaporation which decreased
the film roughness down to 1.4 nm and suppressed the interdiffusion at the
Pb/Cu interface \cite{Neerinck90}. After the lift-off processing, the
samples were characterized by X-ray, AFM and SEM. The X-ray results show
that the Pb/Cu bilayers are polycrystalline with a preferential growth of
the Pb and Cu in the (111) direction. An AFM picture of the 2$\times $2
antidot cluster of Pb/Cu with four leads attached for electrical connection,
is shown in Fig.~\ref{fig:afm}a. Also shown (Fig.~\ref{fig:afm}b) is the AFM
image of a reference sample which consists of a Pb(50 nm)/Cu(17 nm)
microsquare of 2$\times $2 $\mu m^2$ without antidots.

The top Cu layer of 17 nm, evaporated onto the Pb film, was used to protect
the Pb film from oxidation and to enable electrical connection to the
experimental apparatus using a wire bonding technique through the 150$\times
$150 $\mu m^2$ electrical pads of the sample. Besides that, the Cu layer
changes also the coherence length and the penetration depth of the Pb layer.
We discuss the influence of this Cu layer on the superconducting properties
of the Pb/Cu bilayers in terms of proximity induced superconductivity in the
next section.

The electrical transport properties were measured in a $^3$He cryostat using
the four probe technique. The magnetic field was applied perpendicular to
the film surface and the temperature stabilization was better than 0.4 mK.
Since the mesoscopic samples are very sensitive and easily destroyed by
electrostatic charges, all necessary grounding precautions were taken and
1.2 k$\Omega$ resistors were connected in series with each lead to damp
parasitic voltage peaks. The resistance measurements where performed using
an ac-resistance bridge whereas the $V(I)$ characteristics and critical
current measurements where carried out using a dc-power supply and a
nanovoltmeter. The superconducting/normal (S/N) phase boundary was
determined automatically by keeping the sample resistance at a specific
fixed value (usually at 10 \% of the normal state resistance) and varying
the magnetic field and temperature. Similarly, the critical current
measurements as function of field were performed by keeping the sample at a
certain fixed voltage and varying the applied current and magnetic field.

\subsection{Superconducting parameters of the Pb/Cu bilayer}

\label{sec:suppar}

In order to study the Pb(50 nm)/Cu(17 nm) bilayer which was used for the 2$%
\times $2 antidot clusters, we prepared also a single Pb(50 nm) and Cu(17
nm) film and measured their properties. Below, we present the summary of
these results.

Comparing the resistivity due to electron-phonon scattering, for the Pb(50
nm) film, $\rho _{300K}-\rho _{7K}$ = 27.8 $\mu \Omega cm$ and for the $17$
nm Cu thin films $\rho _{300K}-\rho _{7K}$ = 12.47 $\mu \Omega cm$, with the
published bulk values of \mbox{21 $\mu \Omega cm$} and
\mbox{1.7
$\mu \Omega cm$} respectively \cite{Meaden65}, we conclude that the
resistivity of the Pb film approaches the bulk value whereas the resistivity
of the Cu film is enhanced with respect to the bulk value due to the finite
size effect \cite{Neerinck91}. The mean free paths at 7K determined from the
resistivity values are $l_{Pb}$ = 33 nm and $l_{Cu}$ = 7.2 nm respectively.
Thus, both single films, Pb (50 nm) and Cu (17 nm) are in the dirty limit
since $l_{Pb}<\xi _0$ and $\lambda _{Cu}>>1$ \cite{Biagi85} respectively,
where $\xi _0$ = 83 nm is the BCS coherence length of the Pb \cite
{DeGennesbook} and $\lambda _{Cu}=({\hbar v_F})/({2\pi Tl_{Cu}k_B})=38$ at 7
K is a dimensionless impurity parameter.

The superconducting properties of the S layer are characterized by two
important lengths, the temperature dependent coherence length, $\xi
_S(T)=0.85\frac{\sqrt{\xi _0l_{Pb}}}{\sqrt{{1-T/{T_c}}}}$ and the
penetration depth, $\lambda _S(T)$ =$0.66\lambda _L\frac{\sqrt{{\ \xi _0}/{\
l_{Pb}}}}{\sqrt{{1-T/{T_c}}}}$ in the dirty limit. Here $\lambda _L$ is the
London penetration depth \onlinecite{DeGennesbook}. For the Pb (50 nm) film
at $T$= 0 K, these two quantities take the values $\xi _{Pb}(0)$ = 45 nm and
$\lambda _{Pb}(0)$ = 39 nm, respectively. Experimentally, we can also
determine $\xi _{Pb}(0)$ from the relation between the perpendicular second
critical field $B_{c2,S}$ in a S layer and its in-plane superconducting
coherence length, $\xi _S(0)$, given by
\begin{equation}
B_{c2,S}(0)=\frac{\Phi _0}{2\pi {\xi _S}(0)^2}  \label{Eq:perpB}
\end{equation}
where $\Phi _{\circ }=h/2e$ is the flux quantum. The coherence length
obtained in this way is $\xi _{Pb}(0)$=36 nm. Now, we can determine the
Ginzburg-Landau parameter $\kappa _{Pb}={\ \lambda }_{Pb}{(0)}/{\xi }_{Pb}{%
(0)}=1.08$ and since $\kappa _{Pb}>0.7$ we conclude that the single Pb layer
is a type-II superconductor.

By covering the Pb with a Cu layer, a superconducting Pb/Cu bilayer is
obtained with different characteristic parameters than those of the single
Pb film. We therefore define the effective superconducting parameters as the
ones that can be determined from measurements on a Pb/Cu bilayer.

The Pb/Cu antidot clusters had a superconducting transition temperature, $%
T_{c_{NS}}=6.05$K, whereas the reference sample (i.e. square of 2$\times $2 $%
\mu m^2$ without antidots) had a $T_{c_{NS}}=5.55$K. The systematic
difference in critical temperature between the perforated and reference
sample is probably related to a difference of the electrical properties of
the interface between the Pb and Cu layers. The possibility for an oxidation
at the Pb/Cu interface is higher in perforated samples. This results in a
smaller proximity coupling and, possibly, a higher effective critical
temperature. Nevertheless, it is obvious that in both cases the top Cu layer
decreases $T_{c_{NS}}$ below the $T_c=7.2$K of the bulk Pb due to the
proximity effect \cite{DeGennes64}.

From the measured critical fields of the Pb(50 nm)/Cu(17 nm) bilayer, we
determined an effective superconducting coherence length as $\xi _{Pb/Cu}
(0) $=39 nm, using Eq.~\ref{Eq:perpB}.

One may expect that an effective penetration depth for the (NS) bilayer, $%
\lambda _{NS}(0)$ could also be defined. Especially, since there should
exist a penetration depth, $\lambda _N(0)$, determining the decay of the
magnetic field in the normal layer with proximity coupling, which should be
different from $\lambda _S(0)$. The exact definition of these two
quantities, $\lambda _N(0)$ and $\lambda _{NS}(0)$, is not obvious. Since we
know that $\lambda _{NS}(0)>\lambda _S(0)$, we will estimate a lower limit
for $\lambda _{NS}(0)$ as $\lambda _{Pb}(0)=39$ nm. Further in the paper
(see section~\ref{sec:IV}), a more accurate estimate for $\lambda _{NS}(0)$
will be given.

With this information, we can determine a lower limit for the effective
Ginzburg-Landau parameter $\kappa _{Pb/Cu}={\lambda _{Pb/Cu}(0)}$/${\xi
_{Pb/Cu}(0)}$ as 1 for the Pb(50 nm)/Cu(17 nm) bilayer. Since $\kappa
_{Pb/Cu}>0.7$ the Pb/Cu bilayer is a type-II superconductor.

As shown above, by covering Pb with a thin Cu layer, we are not only
protecting the former against oxidation, but we are also changing, in
certain limits, the effective $\xi $ and $\lambda $.

Pb/Cu bilayers are good candidates for studying proximity coupling effects
\cite{Gilabert77} between a superconductor (S) and a normal metal (N) since
interdiffusion hardly exists at the interface of these materials, especially
when they are evaporated at low temperature. Therefore, each material
remains within its own boundaries.

The proximity effect \onlinecite{DeGennes64} refers to the induction of
superconductivity in a contacting normal metal by the decay of the
superconducting order parameter, $\Delta $, from a superconductor into a
normal metal. Not only the order parameter in the superconductor decays when
approaching the S/N interface from the superconducting size but also a non
vanishing superconducting order parameter nucleates in the N layer close to
the interface. Thus, one of the most important parameters defining the
proximity effect between two layers is the so-called coherence length of the
normal layer, $\xi _N$, which determines the length in the normal metal over
which Cooper pairs can diffuse (see Fig~\ref{fig:prox}). In the dirty limit
and for our Cu parameters, $\xi _{Cu}={(\hbar v_Fl_{Cu}}/{6\pi k_BT)}^{1/2}$
= 28 nm at 6 K, where $v_F$ is the Fermi velocity of Cu. Note that $\xi
_{Cu}>d_{Cu}$ where $d_{Cu}$ = 17 nm is the Cu thickness, which means that
the superconducting order parameter is finite over the full thickness of the
Cu layer in the temperature interval at which experiments on the Pb/Cu
antidot cluster were performed (5 K $<\,T\,<$ 6 K).

Another very interesting parameter in proximity induced superconductivity is
the extrapolation length \onlinecite {Gilabert77}, $b=\gamma \xi
_{Cu}(T)\coth ({d_{Cu}}/{\xi (T)})$ = 98 nm at 6 K in the dirty limit (see
Fig.~\ref{fig:prox}), where $\gamma ={\rho _{Cu}}/{\rho _{Pb}}$. This
parameter $b$ is a measure of the unfavourable influence of the N layer on
the superconductivity of the S layer. If $b<<\xi _S(T)$ the effects of
proximity coupling are important whereas if $b>>\xi _S(T)$ the
superconducting order parameter at the interface is almost not changed and
the effects of proximity coupling are negligible. In our case at $6$K, $b$ =
98 nm, $\xi _{Pb}(6K)$ = 108 nm, thus the proximity effect has an
intermediate strength in the Pb(50 nm)/Cu(17 nm) bilayer.

We can classify our samples as being in the Cooper limit %
\onlinecite{Gilabert77} since they fulfill the constraints $d_{Cu}<\xi _{Cu}$
and $d_{Pb}<\xi _0$. In this limit, $\Delta (x)$ can be taken as a constant
over the individual layers and discontinuous at the interface. The
perpendicular second critical field, $B_{c2,NS}$ for a (NS) bilayer of
thickness ($d_N+d_S$) can then be determined from the second critical field
of the single superconducting layer, $B_{c2,S}$, by the dirty limit
expression \onlinecite {Biagi85},
\begin{equation}
B_{c2,NS}(0)=\frac{B_{c2,S}(0)}{1+\eta {d_N/d_S}}  \label{Eq:Bc2}
\end{equation}
where $\eta =\rho _S/\rho _N$ for a specular scattering at the interface.
This expression is in good agreement with our experimental results on
bilayers with different Cu thicknesses. In particular, for the bilayer Pb(50
nm)/Cu(17 nm) used for the fabrication of the antidot cluster, Eq.\ref
{Eq:Bc2} reduces to: $B_{c2,NS}(0)=\alpha B_{c2,S}(0)$ with $\alpha $=0.91.
Experimentally, we have obtained $\alpha $=0.86 which is in agreement with
the calculated value within a 5\%. Taking into account that the scattering
at the interface is probably not completely specular, the obtained agreement
is quite good.

Combining Eq.~\ref{Eq:perpB} and Eq.~\ref{Eq:Bc2}, one can obtain the ratio
between the superconducting coherence length for the Pb(50 nm) layer and
that of the Pb(50 nm)/Cu(17 nm) bilayer, ${\xi _{Pb/Cu}(0)}/{\xi _{Pb}(0)}$=
1.05 . Using the value $\xi _S(0)$= 36 nm for the Pb(50 nm) film, we expect,
according to this relation, that $\xi _{Pb/Cu}=1.05\;\xi _S(0)$ = 38 nm for
the Pb(50 nm)/Cu(17 nm) bilayer film, which is within 5\% agreement with the
'experimental' value determined from the measured second critical fields of
bilayers of that specific thickness, $\xi _{Pb/Cu}(0)$=39 nm .

In summary, from the analysis of the superconducting properties of the Pb/Cu
bilayer we can conclude that: i) The Pb(50 nm)/Cu(17 nm) bilayer shows
proximity induced superconductivity of an intermediate strength. ii) The
proximity induced superconductivity penetrates through the whole thickness
of the Cu layer. iii) The effective parameters, $B_{c2,NS}(0)$, $\xi
_{NS}(0) $ and $\lambda _{NS}(0)$ for the bilayer films have been determined
from the experimental results.


\section{Results and discussion}

\label{sec:resdisc}

\subsection{The S/N phase boundary}

\label{sec:phasbo}

\subsubsection{Experimental results}

\label{sec:phasboexp}

Fig.~\ref{fig:psbexp1} shows the (S/N) phase boundary, $\Delta T_c(B)$=$%
T_c(0)-T_c(B)$, of the Pb(50 nm)/Cu(17 nm) 2$\times $2 antidot cluster
measured with the criterium of 10\% of the normal state resistance.
Pronounced periodic oscillations of $T_c(B)$ are observed every 26 G.
Defining a flux quantum per antidot as $\Phi _{\circ }=h/2e=B\cdot S$, where
$S$ is an effective area per antidot ($S=0.8\mu m^2$), the oscillations with
periodicity of 26 G can be correlated with a magnetic flux per antidot, $%
\Phi =n\Phi _{\circ }$, where $n$ is an integer number. In each of these 26
G periods, smaller dips appear at approximately 7.5 G, 13 G and 18 G. The
smaller dips correspond to approximately $\Phi /\Phi _{\circ }=0.3$, 0.5 and
0.7.

Superimposed with these oscillations, a parabolic background is observed,
reflecting the $B(T)$ dependence of the second critical field of the
quasi-one-dimensional stripes constituting the antidot cluster, $%
B_{c2}^{1D}=({\sqrt{12}\Phi _0})/({2\pi w\xi (T)})$ \cite{Tinkham75} where $%
w $ is the width of the stripe. From this formula, an effective coherence
length $\xi _{NS}(0)$= 35 nm is obtained for this particular sample with $%
T_c $ =6.05 K. Above 52 G, deviations from the main 26 G periodicity and
intermediate substructure are observed.

In the inset of Fig.~\ref{fig:psbexp1}, two magnetoresistance curves, $R(B)$%
, measured at $T_1$ = 6.009 K and $T_2$= 5.972 K are shown. At $T_1$ only
minima at $n\times $ 26 G are distinctly observed. At $T_2$ an additional
substructure appears in each period. The magnetoresistance results confirm
the features observed in the $T_c(B)$. A cut of the $T_c(B)$ boundary at
high temperature (i.e. $T\sim T_1$) reveals only the main 26 G oscillations,
whereas a cut at lower temperatures results in the full substructure.

Fig.~\ref{fig:psbref} shows the (S/N) phase boundary, $\Delta T_c(B)$=$%
T_c(0)-T_c(B)$, of the reference sample (i.e. a superconducting microsquare
without antidots (see Fig.~\ref{fig:afm}b)) measured at 10\% of the normal
state resistance. In this case, oscillations are also observed however they
are clearly different from those presented in Fig.~\ref{fig:psbexp1}. The
first oscillation of $\Delta B_1\sim $ 16 G is followed by quasi periodic
oscillations with decreasing $\Delta B$ $(\Delta B_2=9.9\;G,\;\Delta
B_3=8.2\;G,\;\Delta B_4=7.5\;G,\;\Delta B=7.1\;G)$ In addition, the
superimposed background of Fig.~\ref{fig:psbref} follows a linear relation
instead of the parabolic dependence seen in Fig.~\ref{fig:psbexp1}. The
oscillations are caused by the confinement of the superconducting condensate
in the dot as revealed by other studies \cite{Moshch95}\cite{Benoist96}. The
remark must be made that the first two periods are noticably larger than
predicted in Ref.\cite{Benoist96}.The other periods correspond to the
calculations within 15\% The linear background observed in Fig.~\ref
{fig:psbref} is related to the second critical field of a 2D system, $%
B_{c2}^{2D}$, is given by the expression $B_{c2}^{2D}={\Phi _0}/{2\pi \xi
(T)^2}$. From this linear background an effective coherence length, $\xi
_{NS}(0)$= 41 nm is determined for this specific sample with $T_c$= 5.55 K.

In the inset of Fig.~\ref{fig:psbref}, the magnetoresistance curves for the
reference sample are shown at four different temperatures. For any of these $%
R(B)$ curves, two large drops of resistance are observed when decreasing the
magnetic field. In addition, for each curve and at low fields, one single
small dip appears. This dip shifts to higher magnetic fields when decreasing
the temperature and is responsible for the oscillations observed in $T_c(B)$%
. On the contrary, the two drops of resistance may be explained by the
influence of the four narrow superconducting leads which are attached to the
microsquare (two for sending the transport current and two to measure the
voltage). These leads are quasi-one-dimensional wires and therefore they
become superconducting below the field
\mbox{$B_{c2}^{1D}=({\sqrt{12}\Phi
_0})/{2\pi w \xi(T)}$}. However, the dot itself has a second critical field
transition given by \mbox{$B_{c2}^{2D}={\Phi _0}/({2 \pi\xi(T)^2})$}, as
mentioned above.

Taking into account that the transition to the superconducting state takes
place at higher magnetic fields for the 1D system (i.e. leads) than for the
2D system (i.e. microsquare), one expects to attain the following situation:
when decreasing the magnetic field, starting from the normal state, a first
resistive drop is observed, caused by a transition to the superconducting
state of the narrow current and voltage leads. The fact that the resistance
of the microsquare is affected by this transition, does not mean that the
''four point'' resistance measurement contains a portion of the leads, but
arises from those parts of the cluster which become superconducting due to
their proximity to the leads. The theoretical critical field values for the
leads, as calculated from the the above formula for $B_{c2}^{1D}$ using the
effective coherence length $\xi _{NS}$ obtained from the linear background
of $T_c(B)$, are marked by the open squares in the inset of Fig.~\ref
{fig:psbref}. The first resistance drop of the dot takes place at slightly
lower fields than the N/S transition of the leads, since they are only in
proximity coupling with the dot. Therefore, the squares indicating $%
B_{c2}^{1D}$ fall on a horizontal straight line above the onset of the
resistance drop, and not in the middle of the transition.

Further decreasing the magnetic field, we cross the line $B_{c2}^{2D}$ at
which the microsquare becomes superconducting and a second resistance drop
is initiated (see open circles in the inset of Fig.~\ref{fig:psbref}). In
this part of the curve, small dips appear in the magnetoresistance due to
the dot geometry. These dips are responsible for the oscillations observed
in the $T_c(B)$ curve. Similar two-step $R(B)$ transitions have been
observed by Chi $etal.$ \cite{Chi94}, who studied narrow 1D superconducting
wires connected to large 2D contact pads. In their case, the proximity
coupling between the 1D wires with a large critical field and the 2D pads
with a small critical field gave rise to a two-step behaviour comparable to
our results.

We have to note that the expression used to calculate the $B_{c2}^{2D}$
values corresponds only to the linear $T_c(B)$ background; the periodic
oscillations superimposed with it are not considered in this expression and
thus, it is not surprising that the calculated values for $B_{c2}^{2D}$ (see
open circles in the inset of Fig.~\ref{fig:psbref}) do not lie on one
horizontal line. However, if we determine the critical magnetic field values
from the measurements of $T_c(B)$ (see Fig. 4) where both the background and
oscillations due to the dot geometry are considered at a criterium of 10\%,
we obtain the open triangles shown in the inset. Their position agrees with
all the expectations, i.e. the triangles lie on a horizontal line which is
at 10\% of the normal state resistance.


\subsubsection{Calculations of the Tc(B) phase boundary}

\label{sec:phasbotheo} In order to calculate the S/N phase boundary of the 2$%
\times $2 antidot cluster, we have approximated the cluster geometry by a
square network of quasi-one-dimensional superconducting wires with a width $%
w $ and a length $l$ (see Fig. \ref{fig:cluster}). Adjacent nodes $(i,j)$
are coupled by a supercurrent depending on the gauge-invariant phase
difference $\gamma _{ij}$ between the nodes~:
\begin{equation}
\gamma _{ij}=\phi _j-\phi _i+\frac{2\pi }{\Phi _{\circ }}\int_i^j{\bf A}%
\cdot d{\bf l}  \label{eq:gauginvphas}
\end{equation}
where $\phi _i$ is the phase of the superconducting order parameter at site $%
i$, $\Phi _{\circ }\,=\,h/2e$ the flux quantum, ${\bf A}$ the magnetic
vector potential and $d{\bf l}$ a segment along the wire. First of all, we
have used the approximation developed for weakly coupled wire networks \cite
{Giroud92,Vanderzant94,Chi92,Puig96}, the ''interacting loop-current (ICL)
model'', which neglects the variation of the order parameter along the
superconducting wires of the network and only considers the phase
variations. This simplification is therefore equivalent to the London limit
of the Ginzburg-Landau theory, and leads to a linear current-phase
relation~:
\begin{equation}
I_{ij}=\frac{\Phi _{\circ }wd}{2\pi \mu _{\circ }\lambda ^2l}\gamma
_{ij}=I_{\circ }\gamma _{ij}  \label{eq:supercur}
\end{equation}
where $I_{ij}$ is the supercurrent from $i$ to $j$, $d$ is the wire
thickness and $\lambda $ is the superconducting penetration depth of the
material. The resulting kinetic energy for each wire $(i,j)$ can be
expressed as~:
\begin{equation}
E_{ij}=\frac{\Phi _{\circ }^2wd}{4\pi ^2\mu _{\circ }\lambda ^2l}\gamma
_{ij}^2  \label{eq:nodenergy}
\end{equation}
and is a quadratic function of phase differences. The prefactor in Eq.~\ref
{eq:nodenergy} defines the coupling strength between the adjacent nodes and
is denoted as~:
\begin{equation}
E_J\equiv \frac{\Phi _{\circ }^2wd}{4\pi ^2\mu _{\circ }\lambda ^2l}=\frac
\hbar {2e}I_{\circ }  \label{eq:couplingenergy}
\end{equation}
To obtain the possible energy states of the cluster, all phase differences $%
\gamma _{ij}$ have to be determined. This is done by imposing current
conservation at each node~:
\begin{equation}
\sum_jI_{ij}=I_i^{ext}  \label{eq:curcons}
\end{equation}
where $I_i^{ext}$ is the transport current fed into node ($i$), and by
applying the fluxoid quantization condition to each individual loop :
\begin{equation}
\sum_{(i,j)}\gamma _{ij}=2\pi (n_k-\frac \Phi {\Phi _{\circ }})
\label{eq:fluxquant}
\end{equation}
where the sum is taken over the wires of the $k^{th}$ cell, $n_k$ ($k$ =
1\ldots 4) is the flux quantum number of the $k^{th}$ cell, and $\Phi $ the
magnetic flux. An estimation of the self-inductance for an individual cell
of the antidot cluster gives ${\cal L\approx }$2 pH. Since most of the
results are obtained close to $T_c$, the flux generated by the circulating
currents is negligibly small. As a consequence we have not taken the self-
and mutual inductance effects into account in our theoretical analysis and $%
\Phi $ is considered to be equal to the external magnetic flux. By solving
Eqs.~\ref{eq:curcons} and \ref{eq:fluxquant} for the $\gamma _{ij}$'s at a
given magnetic field and summing the corresponding energies $E_{ij}$ over
all branches ($i,j$) of the system, we obtain the total kinetic energy of
the 2$\times $2 antidot cluster~:
\begin{equation}
E=\frac 12\sum_{(i,j)}E_J\gamma _{ij}^2  \label{eq:totenergy}
\end{equation}
Since the phase differences are found from Eqs.~\ref{eq:curcons} and \ref
{eq:fluxquant}, the total energy $E$ for each set of quantum numbers $n_k$ ($%
k$=1,\dots ,4) becomes dependent on the magnetic flux through the cells. In
Fig.~\ref{fig:psbth1}a the total energy $E$ is plotted as a function of the
magnetic flux for the case of $I_i^{ext}=0$ and for all wires with identical
lengths and widths. We have omitted the parabolic contribution arising from
the finite width of the wires. Since the energy is periodic in $\Phi $ with
a period $\Phi _{\circ }$, we have only plotted the first period. Each set
of $n_k$ ($k$ = 1\ldots 4) values corresponds to a different vortex
configuration and gives rise to an energy branch which is quadratic with the
magnetic flux. Due to the symmetry of the structure, most of the branches
are degenerate and only six principal parabolae can be distinguished. The
S/N phase boundary is given by the branch which has the lowest energy for a
given value of the flux \cite{DeGennesbook}:
\begin{equation}
\Delta T_c(\Phi )=\frac{8\pi ^2\mu _{\circ }\lambda ^2\xi (0)^2T_c}{\Phi
_{\circ }^2V}\,\min_{n_1\ldots n_4}(E)  \label{eq:tcphi}
\end{equation}
where V is the total volume of the structure, $T_c$ is the critical
temperature at zero field and $\xi (0)$ is the coherence length at $T=0$.
The phase boundary is therefore composed of five branches and has three
minima at $\Phi /\Phi _{\circ }=0,\,0.5,\,1$ and four cusps at $\Phi /\Phi
_{\circ }=0.3,0.37,0.63,0.7$ (see Fig.~ \ref{fig:psbth1}a). At the first
branch, around $\Phi =0$, all the $n_k=0$ ($k$ = 1\ldots 4) and a screening
current is present only at the edge of the structure. When $\Phi $
increases, the fourfold degenerate state with one vortex at one of the cells
becomes more stable. The transition occurs at $\Phi /\Phi _{\circ }=0.3$. At
$\Phi /\Phi _{\circ }=0.37$ a doubly degenerate diagonal state becomes
favorable, where two vortices occupy one of the diagonals of the cell. Note
that this state (which is the analog of the ''checkerboard'' vortex
configuration in antidot lattices \cite{VVM}) has a lower energy than the
configuration where two adjacent cells are occupied by vortices (parallel
state). The diagonal state could be of interest for flux quantum logic
applications \onlinecite{Puig96}. In higher fields, a third vortex enters
the cluster at $\Phi /\Phi _{\circ }=0.7$ and finally around $\Phi /\Phi
_{\circ }=1$ all the cells are filled and only an edge current is flowing
now counter clockwise.

It is interesting to note that Eqs.~\ref{eq:supercur} to \ref{eq:totenergy}
are similar to the equations describing the supercurrent and the energy in
Josephson junction arrays \cite{Giroud92}. In JJA's however, the
current-phase relation is sinusoidal and only the phase differences over the
Josephson junctions have to be considered. It turns out that the differences
between the linear and the ''JJA'' models are not very important when the
static properties in a magnetic field are considered. For comparison we show
in Fig.~\ref{fig:psbth2} the $E(\Phi )$ curve for a JJA together with the
one obtained for the linear relation. Nearly identical energy branches are
indeed found caused by the obvious fact that the sinusoidal current-phase
relation can be approximated by a linear one, for small phase differences.

For completeness, we also show in Fig.~\ref{fig:psbth2} the curve predicted
by the linearized de Gennes-Alexander \cite{DeGennesalexander} formalism for
strong-coupling one-dimensional wire networks, which incorporates phase and
amplitude variations of the order parameter along the wires of the network.
This formalism assumes that the diameter of the superconducting strands
which constitute the network is smaller than both the coherence length and
penetration depth. In that way, the problem becomes 1 D, and the order
parameter $\Psi _s$ at a position $s$ on a wire connecting two adjacent
nodes $a$ and $b$ (see Fig.~\ref{fig:cluster}) can be expressed as follows:
\begin{equation}
\Psi _s=\frac{e^{i\gamma _{as}}}{\sin \left( \frac{l_{ab}}{\xi (T)}\right) }%
\left[ \Psi _a\sin \left( \frac{l_{ab}-l_{as}}{\xi (T)}\right) +\Psi
_b\,e^{-i\gamma _{ab}}\sin (\frac{l_{as}}{\xi (T)})\right]  \label{Eq.psidGA}
\end{equation}
where $\Psi _a$ and $\Psi _b$ are the order parameters at the nodes $a$ and $%
b$, $l_{ab}$ the distance between points $a$ and $b$ and $\gamma _{ab}$ the
line integral of the vector potential multiplied by $2\pi /\Phi _{\circ }$.
From the second GL equation it is possible to determine the supercurrent
density through the branch. It depends sinusoidally on the phase difference
between the nodes $a$ and $b$, similar to the current through a Josephson
junction:
\begin{equation}
J_{ab}=\frac{2e\hbar }{m^{*}\xi (T)}\left| \Psi _a\right| \left| \Psi
_b\right| \frac{\sin (\phi _a-\phi _b-\gamma _{ab})}{\sin \left( \frac{l_{ab}%
}{\xi (T)}\right) }
\end{equation}
Applying the standard boundary conditions at a node $a$, gives rise to the
Alexander node equations:
\begin{equation}
\sum_n\left[ -\Psi _a\cot \left( \frac{l_{an}}{\xi (T)}\right) +\Psi _n\frac{%
e^{-i\gamma _{an}}}{\sin \left( \frac{l_{an}}{\xi (T)}\right) }\right] =0
\label{Eq:nodeq}
\end{equation}
which express the current conservation at node $a$ similar to Kirchhoff's
current law for resistor networks. The sum is taken over all nodes $n$ which
are nearest neighbours of node $a$ (see Fig.~\ref{fig:cluster}). If this
equation is written for all nodes in the network, this leads to a
characteristic determinant which must be zero for the existence of a
non-trivial solution. We obtain

\begin{eqnarray}
&&\left( 1+2\cos \frac{2\pi \Phi }{\Phi _0}-3\cos \frac{2l_{ab}}{\xi (T)}%
\right) \cdot \left( 1+2\cos \frac{2\pi \Phi }{\Phi _0}+3\cos \frac{2l_{ab}}{%
\xi (T)}\right) \cdot \\
&&\left( 1+3\cos \frac{2l_{ab}}{\xi (T)}-2\sin \frac{2\pi \Phi }{\Phi _0}%
\right) \cdot \left( 1+3\cos \frac{2l_{ab}}{\xi (T)}+2\sin \frac{2\pi \Phi }{%
\Phi _0}\right)  \nonumber \\
&=&0  \nonumber
\end{eqnarray}
where $\Phi $ is the flux of the applied magnetic field per antidot.\
Equating the first factor to zero, one obtaines a curve defining the first
and the fifth branch of the $T_c(B)$ phase boundary. Likewise, the middle
three branches are determined by the other three factors (see Fig.\ref
{fig:psbth2}).

This linearized approach is valid only close to $T_c$. At lower
temperatures, the non-linear version of this model should be applied \cite
{Fink91a,Fink91b}.

The qualitative agreement between the ILC model (see Fig.~\ref{fig:psbth1}$%
(a)$) and the experimental curves of Figs.~\ref{fig:psbth1}($b$) and ($c$),
where the parabolic background has been subtracted, is fairly good. The
measurement of curve ($b$) was performed with an ac current of 1 $\mu $A,
the one in ($c$) with 3 $\mu $A. The five parabolae corresponding to the
different states are found back in the experimental plots. An important
difference is however, that two maxima appear in the experimental phase
boundaries at $\Phi /\Phi _{\circ }\simeq $ 0.2 and 0.8 which are not
reproduced in the calculations. The $\Delta T_c(\Phi )$ dependence for the
states with all $n_k=0$ or all $n_k=1$ thus seems steeper than predicted by
the theoretical model and it also has a larger amplitude. Strangely, the
agreement with the theoretical curve, which is calculated for zero transport
current, is better in curve ($c$), where the highest transport current was
used.

In addition, the experimental $\Delta T_c(\Phi )$ (see Fig.~\ref{fig:psbexp1}%
) shows a considerable parabolic background, some distortions and a
disappearance of the substructure above $2\Phi _{\circ }$.

As it was mentioned previously, the parabolic background (''diamagnetic
shift'') is due to the penetration of the magnetic field in the volume of
the wires \cite{Pannetier91,Tinkham75} and can be taken into account by
adding the term~:
\begin{equation}
\frac 12\sum_{(i,j)}E_J\frac{\pi ^2B^2l^2w^2}{12\Phi _{\circ }^2}
\label{eq:fieldenergy}
\end{equation}
to the right hand side of Eq.~\ref{eq:totenergy}. The effect of this
background is shown in Fig.~\ref{fig:psbdis}. Curve~$(a)$ shows $\Delta
T_c(\Phi )$ obtained from Eq.~\ref{eq:totenergy} and Eq.~\ref{eq:tcphi} for
the antidot cluster. Curve~$(b)$ considers the case of a finite linewidth
(Eq.~\ref{eq:fieldenergy}) with $w=0.35\,\mu m$ for wires at the edges of
the network and $w=0.4\,\mu m$ for the wires connecting the edges with the
central node (see Fig.~\ref{fig:cluster}). Note in curve ($b$) that as the
field is increased, the $\Delta T_c$ amplitude due to fluxoid quantization
in the structure becomes quite small with respect to the shift caused by the
background and the amplitude of the oscillations gradually decreases.

After discussing in detail the first period of the phase boundary $T_c$($%
\Phi $/$\Phi _0$), we would like to analyze briefly the possible reasons for
the shift of the $T_c$($\Phi $/$\Phi _0$) minima in the higher periods of
the S/N phase boundary, i.e. for 1$<\Phi $/$\Phi _0<$2, etc. (Fig.~\ref
{fig:psbexp1}). Taking into account the ''softness'' of the current loops in
a real structure, which, strictly speaking, is not a one dimensional
network, the presence of disorder of these loops is quite probable.
Therefore, one may assume that one of the reasons of the variation of the $%
T_c$($\Phi $/$\Phi _0$) peaks from period to period can be related to an
areal disorder of ''soft'' current loops in the 2$\times $2 antidot cluster.

Another important factor is the disorder arising from width inhomogeneity,
structural defects, non identical electrical properties of each wire, which
can lead to a modification of the critical current of the individual wires
and to a distribution of the effective area per cell. The importance of
areal disorder was already stressed in studies of the phase boundary in wire
networks \cite{Itzler90,Benz88} where it was shown that areal disorder can
lead to a decay of the oscillation amplitudes and in some cases even to
beatings in the envelope. Measurements on Josephson junction clusters
(JJC's) \cite{Harlingen93} showed that coupling disorder coming from an
unavoidable spread in junction parameters is not averaged out as in large
arrays, leaving a clear trace in the transport properties.

Curve~$(c)$ in Fig.~\ref{fig:psbdis} shows the influence of areal disorder
on the theoretical phase boundary of the antidot cluster. The areal disorder
was introduced by allowing the coordinates of the nodes to vary randomly
within a circle of radius 0.1 $\mu m$ around the node position for the
ordered network. In this way a random distribution of lengths $l_{ij}$ and
cell areas is generated which lifts the degeneracy of the possible states
and changes the relative positions of the different parabolic energy
branches. For small fields (i.e. the first period), the deviations with
respect to curve $(b)$ are not very pronounced and it is still possible to
identify all five parabolae forming the $T_c(B)$ phase boundary. As the
field increases, the oscillation amplitude gets smaller and the positions of
the different branches shift, making the identification of the states less
straightforward. From this analysis it is clear that disorder can indeed
cause a shift of the peaks with increasing magnetic field, as it is observed
experimentally. Finally, in curve ($d$) we have included the contributions
of the parabolic background (curve ($b$)) and the areal disorder (curve ($c$%
)). A comparison of curve ($d$) with the experimental data of Fig.~\ref
{fig:psbexp1} shows that certainly these two contributions are playing a
role in our measurements. Nevertheless, even by including these two effects
in the model, it is still not possible to simulate the experimental curves
completely. This is probably due to the fact that we have used a
one-dimensional approach, which cannot take into account the 2D character of
the structure.

\subsection{Critical currents and $V(I)$ characteristics}

\label{sec:critcur}

So far we have considered the effects of the vortex confinement by the 2$%
\times $2 antidot cluster on the S/N phase boundary $T_c(B)$. In order to
demonstrate that the unique properties of the Pb/Cu 2$\times $2 antidot
cluster are not restricted to the S/N phase transition, we present below the
critical current results and the $V(I)$ characteristics measured at
temperatures \mbox{400 mK $< T_c-T <$ 100 mK}. We will show that at these
temperatures the quantized states are still present and that the most stable
ones at the S/N phase boundary do not always correspond to the states
carrying the highest currents. In addition, we will demonstrate that a
transport current $j\neq $ 0 is able to lift some of the degeneracies of the
vortex configurations at $j$ = 0. The $V(I)$ characteristics will be used to
give a qualitative picture of the flux line transport in these
nanostructures in terms of phase-slip processes.

It is known that the dynamics of 1D wires \cite{Skocpol,Kopnin84} and arrays
of 1D wires \cite{Giroud92} are mostly governed by phase-slip processes when
currents close to the depairing critical current are sent through the wires.
Thus, one may also expect that signatures of these processes should also be
present in the $V(I)$ characteristics of the 2$\times $2 antidot cluster. In
such a phase-slip process the energy of the system is reduced by bringing a
small spot of the 1D wire (of the order of the quasiparticle diffusion
length, $\lambda _Q^{*}\simeq \xi (T)$) momentarily to the normal state.
During this process, the phase of the superconducting order parameter in
that spot is changed by 2$\pi $. To preserve the superconductivity in the
sample in the presence of a large current, the phase-slip process repeats in
time and the average period $\tau $ between phase-slips is related to the
voltage measured between the two ends of the wire through the Josephson
relation, $V=\frac \hbar {{2e}}\frac{{2\pi }}\tau $. The points where the
order parameter becomes zero and its phase shows jumps of 2$\pi $, are known
as phase-slip centres (PSC). Obviously, if the wire already has some weak
superconducting points, the PSC's will be localized at these spots.

Several mechanisms of PSC formation have been reported. First, PSC's can
nucleate when the current exceeds the depairing critical current of the 1D
wire, as described by the theory of Skocpol, Beasley and Tinkham (SBT)\cite
{Skocpol}. In this case the formation of 1,2,\ldots n PSC's gives rise to a
step-like $V(I)$ characteristic.

PSC's can also be formed as a result of thermodynamic fluctuations which
take place with a probability proportional to $\exp (-\delta {\cal F}/k_BT)$%
, where $\delta {\cal F}=w\sqrt{2}E_Jl/3\xi (T)$ is the free energy barrier
between the state before and after the phase-slip. The theoretical
description of PSC's was developed by Langer and Ambegoakar (LA) \cite
{Langer}. Taking into account the exponential decrease of the phase-slip
probability with temperature, their model is only applicable in a very
narrow temperature interval near $T_c$.

A third mechanism has recently been reported by Giroud {\it et al.} \cite
{Giroud92} for arrays of 1D wires with localized vortices. They consider the
possibility that when a current close to the depairing critical current is
reached in one of the wires, the vortex feels a ''Lorentz-like'' force
perpendicular to the transport current, which tends to move the vortex to
the next cell where the process is repeated. This model assumes the
nucleation of a phase-slip in a 1D wire each time that a vortex crosses the
1D wire.

\subsubsection{Critical currents: experimental results and comparison with
the model}

\label{sec:critcurexp}

The critical current density versus magnetic field, $J_c(B)$, of the 2$%
\times $2 antidot cluster was determined with a criterium of 3$\mu V$ as
explained in section~\ref{sec:prepchar}. Then, at specific values of the
magnetic field and temperature, the $V(I)$ characteristics were measured to
confirm the $J_c(B)$ results and study the vortex dynamics in the antidot
cluster.

Fig.~\ref{fig:JcB} shows the $J_c(B)$ curves measured at four temperatures.
Note that clear maxima are observed at $n\Phi _0$ and $(n+0.5)\Phi _0$, and
smaller inflections are detected around $(n+0.3)\Phi _0$ and $(n+0.7)\Phi _0$
(see Fig.~\ref{fig:JcB}b and \ref{fig:JcB}c). Initially, the magnitude of
the oscillations increases by decreasing the temperature (Fig.~\ref{fig:JcB}%
a,b) but they disappear almost completely when further decreasing the
temperature (see Fig.~\ref{fig:JcB}d). A parabolic background caused by the
magnetic field penetration in the elementary wires is again observed at all
temperatures.

The magnetic field values at which specific features appear in $J_c(B)$
(Fig.~\ref{fig:JcB})and $T_c(B)$ (Fig.~\ref{fig:psbexp1}) are the same. Thus
one may assume that the maxima observed in the $J_c(B)$ of the 2$\times$2
antidot cluster are related to certain stable vortex states revealed in the $%
T_c$($B$) oscillations. However the relative amplitude of these maxima in
one period of $\Phi _0$ is different for $J_c(B)$ than for $T_c(B)$. In $%
T_c(B)$ large minima were observed at $n\Phi _0$ and three smaller minima of
the same energy were obtained at around $(n+0.3)\Phi _0$, $(n+0.5)\Phi _0$
and $(n+0.7)\Phi _0$. On the contrary, in $J_c(B)$, the maxima at $n\Phi _0$
are followed in magnitude by the maxima at $(n+0.5)\Phi _0$ whereas the
features appearing at $(n+0.3)\Phi _0$ and $(n+0.7)\Phi _0$ have almost
completely lost their amplitude.

To estimate the field dependence of the critical current theoretically and
compare it with the experimental observations we have used the model
described in section ~\ref{sec:phasbotheo} and solved equations \ref
{eq:curcons} and \ref{eq:fluxquant} for the case where the external
transport current enters the structure at node $(a)$ and leaves at node $(b)$
(see Fig.~\ref{fig:cluster}). The current-phase relation, given by Eq.~\ref
{eq:supercur}, has slightly been modified to take into account the order
parameter depression in the strands when the current is close to the
depairing current \cite{Kopnin84}. We have used the relation :
\begin{equation}
I_{ij}=I_{\circ }(\frac{\gamma _{ij}}{\gamma _c})(1-\frac{1}{3}(\frac{\gamma
_{ij}}{\gamma _c})^2)  \label{eq:curphas}
\end{equation}
which is linear for small phase differences and becomes parabolic near $%
\gamma _{ij}=\gamma _c$, where \mbox{$\gamma_{c}\equiv l/(\sqrt 3 \xi)$} is
the critical phase difference at which a phase-slip process occurs and %
\mbox{ $I _{\circ}\gamma_{c}= I_{dep}^{GL}$} is the Ginzburg-Landau
depairing current. Note that Eq.~\ref{eq:curphas} is identical to the
current-phase relation obtained for a long ideal weak link in the depairing
limit \cite{Likharev79}.

The set of nonlinear equations for the unknown phase differences $%
\gamma_{ij} $ obtained from Eqs.~\ref{eq:curcons} and \ref{eq:fluxquant},
was solved numerically using the standard Newton-Raphson method \cite
{Press90}. The critical current at a magnetic flux $\Phi$ and a fixed state $%
n_{k}$ ($k = 1\ldots 4$) was obtained as follows : first, the $\gamma_{ij}$%
's were initialized to zero, $I_{a}^{ext}$ was fixed at a certain value and
the corresponding phases were determined. Next, the external current was
slowly ramped up and the same procedure was repeated until the current was
too high to find any solution for the set of equations. The external
current, above which no static solution exists, has been taken as the
intrinsic critical current of the structure. The self-field induced by the
transport current was estimated to be not higher than 0.1 $\%$ and was
therefore neglected in the analysis.

Curve $(a)$ in Fig.~\ref{fig:critcurtheo} shows the theoretical $I_c(\Phi )$
obtained with Eq. \ref{eq:curphas} and $\gamma _c=\pi /2$. Only the states
with the highest critical current at that particular $\Phi $ are shown and
we have restricted ourselves to the first period, without considering any
disorder. Because of the current injection at node ($a$) the symmetry is
broken and the resulting vortex configurations differ from the ground-state
configurations at zero applied current, discussed in section~\ref
{sec:phasbotheo}.

The possible states are displayed schematically in Figure ~\ref
{fig:critcurtheo}$(a)$. Only two states with one vortex in the structure are
possible, instead of the fourfold degenerate ground-state obtained for the
case of $I_a^{ext}=0$. Near $\Phi =\frac 12\Phi _{\circ }$ the supercurrent
is carried by a state where the vortices occupy the second row of the
cluster (parallel state) instead of being located on the diagonals
(checkerboard configuration).

The reason for the substantial modification of the stable vortex
configurations is of course that the external current is added to the
circular currents which flow to satisfy the fluxoid quantization. The total
amount of current that can be injected before the structure depairs is
higher if the transport and circular shielding currents are subtracted. For
the case of large weakly-coupled wire networks \cite{Vanderzant94} and
inductive JJA's \cite{Phillips94} it was already predicted that large
transport currents make the checkerboard ground-state unstable. This leads
to a state of parallel vortex rows which can move coherently (driven vortex
lattices) over the underlying periodic array.

Curve $(b)$ in Fig.~\ref{fig:critcurtheo} shows the static calculation for a
JJC having a sinusoidal current-phase relation for comparison. Besides some
minor differences concerning the exact positions of minima and maxima
(compare curves $(a)$ and $(b)$ in Fig \ref{fig:critcurtheo}), the same
states can be identified in the $I_c(\Phi )$ plot. Note however, that in
this case the normalizing current $I_{c\circ }$ is the Josephson critical
current and not the depairing current.

If in the case of the antidot cluster (curve ($a$)), $\gamma_c>\pi/2$, the
overall shape of the calculated $I_c(\Phi)$ curve slightly changes and the
difference with the JJC curve is more important, although still the same
states carry the largest supercurrent.

If we compare the theoretical curve ($a$) with the first period of the
experimental data shown without the background for $T$=5.808 K in Fig.~\ref
{fig:critcurtheo}($c$), we note that the curves are qualitatively similar.
The same five states are clearly present. The state at $\Phi /\Phi _{\circ
}=0.5$ has a higher $I_c$ than the states around $\Phi /\Phi _{\circ }=0.3$
and $0.7$ though it does not reach the same current value as predicted in
the calculation, except for temperatures below approximately 5.715 K.

Below this temperature ($T$=5.715 K), the peaks at $n\Phi _{\circ }$ flatten
and the amplitude of the modulations of the critical current decreases with
decreasing temperature. It seems that the intrinsic $I_c(\Phi )$ behaviour
of the structure is cut off by a superimposed parabolic $I_c^{^{\prime
}}(\Phi )$ background (see the full line in Fig.~\ref{fig:JcB}). The
disappearance of the substructure with decreasing temperature was also
observed in wire networks \cite{Vanderzant94} and attributed to an increase
of the energy barrier required to cross the superconducting wires.

In our case however, it looks more like a cut-off rather than a continuous
decrease. Therefore, we believe that the effect could be caused by the
propagation of heat generated in the current leads. Since the antidot
structure can be considered as a kind of parallel circuit having a critical
current which is a factor approximately 1.7 times higher than the depairing
current ( $I_{dep}^{GL}$) of a single wire (see Fig.~\ref{fig:critcurtheo})($%
a$), the current lead is probably in a resistive state while the structure
is still not. We measured the voltage between the nodes $(a)$ and $(b)$ (see
Fig. \ref{fig:cluster}) thus the voltages appearing over the current leads
should not have influenced the results. However, at lower temperatures (when
$I_c$ is rather high), the actual heating due to this current cannot be
excluded. If such heat propagates towards the antidot cluster, it could
trigger a transition to a dissipative state earlier than expected. In that
case, the $I_c(\Phi )$ dependence would show a ''cut-off'' governed by the
parabolic critical current dependence of the current-leads, $I_c^{^{\prime
}}(\Phi )$.

Besides this influence of the current leads at high current levels, care
should also be taken not to overestimate the limits of validity of the
static simulation itself. Unlike the case of JJA's where the dynamics can
relatively easy be taken into account by using the resistively shunted
Josephson junction (RSJ) model in combination with the Josephson voltage
relation \cite{Chung89}, it is not so straightforward to describe the
dynamics in wire networks where resistive PSC's nucleate and can be
activated throughout the structure. The present model does not take into
account any dynamic effects such as vortex motion or propagating PSC's and
therefore, the critical currents obtained by the static simulation should be
considered as an estimate of the upper limit, rather than an accurate
determination.

\subsubsection{$V(I)$ characteristics: results and discussion}

\label{sec:IV}

Fig.~\ref{fig:IV} shows three $V(I)$ curves measured at $T$=5.715 K for
three magnetic fields as representative data. Note that in all the curves,
several steps are observed which depend on the applied magnetic field.
However, an enlargement of the lower part of the $V(I)$ characteristics (see
Fig.~\ref{fig:IV2}) shows that a quasi-linear dissipative foot is clearly
observed for the cases of $\Phi \neq n\Phi _0$. Note that this foot strongly
depends on the magnetic field. In Fig.~\ref{fig:IV2} we show three curves
corresponding to $\Phi =\Phi _0$, $\Phi =0.37\Phi _0$ and $\Phi =0.5\Phi _0$
for $T$ = 5.808 K and $T$=5.715 K. Indicated by a horizontal dotted line is
the 3$\mu $V criterium used to determine the $I_c(\Phi )$ of Fig.~\ref
{fig:JcB}.

When $\Phi =n\Phi _0,$ no foot is observed at any temperature. Fig.~\ref
{fig:IcT} shows the linear dependence of the critical current, $I_c^{2/3}$,
of the Pb/Cu 2$\times $2 antidot cluster versus $T/T_c$ for the measurements
performed at zero magnetic field. $I_c$ follows the $(1-T/T_c)^{3/2}$
dependence expected from the Ginzburg-Landau depairing critical current of
1D wires. The extrapolation of the $(1-T/T_c)^{3/2}$ law to $T/T_c=0$
results in $I_c(0)$=7 mA.

On the other hand, an estimation of the theoretical Ginzburg-Landau value
for the depairing critical current of a 1D wire gives
\begin{equation}
I_{dep}^{GL}(0)=\frac{\Phi _0wd}{3\sqrt{3}\pi \mu _0\lambda (0)^2\xi (0)}
\label{eq:Ic0}
\end{equation}
where $w$ and $d$ are the width and thickness of the wires respectively and $%
\xi (0)$ and $\lambda (0)$ were determined in section~\ref{sec:suppar}. As
shown in the previous section (see Fig.\ref{fig:JcB}($a$)), the depairing
current of the structure is
\mbox{$I_{c}^{th}(0)\approx 1.6
I_{dep}^{GL}(0)=72$ mA}, which is one order of magnitude higher than the
experimental value obtained. It should however be noted that we have
underestimated the value of $\lambda (0)$ by equating it to that of the
single Pb layer, $\lambda _{Pb}(0)$, while, in fact, we should use $\lambda
_{NS}>\lambda _{Pb}$ (see section \ref{sec:suppar}). From the above
analysis, we can estimate an effective penetration depth for the Pb/Cu
bilayer \mbox{$\lambda_{NS}(0) \approx
3\lambda_{Pb}(0)$}.

When $\Phi \neq \Phi _0$, we believe that the field dependent linear foot
which appears at low currents in the $V(I)$ characteristics, is induced by
phase-slip processes. More precisely by the mechanism reported by Giroud $%
{\it et}$ ${\it al.}$\onlinecite{Giroud92} for arrays of 1D wires which
claims that vortices may move from one cell to the other in a direction
perpendicular to the transport current in a kind of stationary ''flux-flow''
regime. In this model a vortex jump from one antidot to another should lead
to a phase-slip induced in the 1D wire crossed by the vortex. Dissipation in
arrays due to the vortex motion has also been reported by other authors \cite
{Vanderzant90,Vanderzant94}. This mechanism may explain the disappearance of
the resistive foot at $\Phi =n\Phi _0$. Since we have shown that at $\Phi
=n\Phi _0$ an edge surface current is flowing in the antidot cluster, no
vortices are present inside the antidot cluster. Thus, if the dissipative
foot is induced by the motion of vortices, no foot should be expected at $%
\Phi =n\Phi _0$.

The possibility of vortex motion was not considered for the critical current
calculation in the previous section. There we determined the critical
current of the structure using a static approach, i.e. the $I_c$ was defined
as the maximum $I$ which could keep zero voltage through the structure, and
still fulfill the fluxoid quantization condition and current conservation
equations. The real critical current values may therefore be lower than the
ones obtained by means of the static approach.

The experimental evidence supporting the idea that motion of vortices in the
antidot cluster induces phase-slip processes and thus gives rise to
dissipation at $I<I_c$, is presented below.


Fig.~\ref{fig:dVdI} shows the differential resistance, dV/dI, as a function
of the reduced flux, $\Phi /\Phi _{\circ }$, determined at the onset of the
dissipative foot in the $V(I)$ characteristic for $T$=5.808 K and $T$=5.715
K (Fig.\ref{fig:IV2}). Since the $V(I)$ characteristic for the dissipative
foot is quasi-linear, we have extracted one single value for the
differential resistance at each $\Phi /\Phi _{\circ }.$ Fig.~ \ref{fig:dVdI}
shows the dV/dI for the $\Phi /\Phi _{\circ }$ interval where the current is
flowing in the internal strands of the 2 $\times $ 2 antidot cluster. The
interval where only an edge surface current is flowing in the antidot
cluster has been omitted since there, no foot is observed. We have denoted
by vertical dashed lines the $\Phi /\Phi _{\circ }$ values at which a change
of vortex configuration was experimentally observed. Note that at both
temperatures, the dV/dI can be defined by two levels: a low level which
accounts for the one-vortex configuration (N=1) and three-vortex
configuration (N=3) or also called one-antivortex configuration, and a high
level given by the two-vortex configuration (N=2). This two level
representation strongly suggests that the dissipative foot might be caused
by a ''flux-flow'' motion of vortices as reported also by Giroud ${\it {et}}$%
${\it {al.}}$ \onlinecite{Giroud92} for wire networks. In this model, a
vortex crossing one cell is associated to a $2\pi $ phase jump, which
generates a voltage in a time $\tau $ given by the Josephson relation,
\begin{equation}
V=\frac \hbar {2e}\frac{2\pi }\tau  \label{eq:joseph}
\end{equation}
If we assume that the stationary dissipation process can be described by a
constant average vortex velocity $v$ (as it the case of classical flux-flow
in type-II superconductors \cite{Bardeen} and JJA's \cite{Orlando91}), the
voltage generated by the net motion of $N$ vortices can be written as~:
\begin{equation}
V=\frac \hbar {2e}\frac{2\pi }\tau f=N\Phi _{\circ }\frac va
\label{eq:josephbis}
\end{equation}
On the other hand, since this model associates a PSC to each vortex jump, we
have equated the total average power $V(I-I_c(\Phi ))$ generated in the
cluster to the sum of powers dissipated by each individual PSC, $%
P_{PSC^{\prime }s}=Ni_Q^2R_n2\lambda _Q^{*}/{a}$, where here $N$ is the
number of PSC's, $i_Q$ is the quasi-particle current flowing in each wire, $%
R_n$ is the normal state resistance of a single wire and $\lambda _Q^{*}$ is
the quasiparticle diffusion length. From this equality and assuming that $%
i_Q\simeq (I-I_c)/3$ \cite{Giroud92,Skocpol}, we have determined an average
vortex velocity,
\begin{equation}
v\approx \frac{R_n2\lambda _Q^{*}(I-I_c)}{9\Phi _{\circ }}
\label{eq:velocity}
\end{equation}
which is proportional to the current $(I-I_c)$. By substituting this
velocity in Eq.~\ref{eq:josephbis} and calculating the differential
resistance at the critical current ($I_c$), we obtain:
\begin{equation}
\left( \frac{dV}{dI}\right) _{I_c}=\frac 29\frac{NR_n\lambda _Q^{*}}a
\label{eq:dVdI}
\end{equation}
From Eq.~\ref{eq:dVdI} and taking $\lambda _Q^{*}\simeq \xi (T)$ and N=1 for
the one-vortex configuration and N=2 for the two-vortex configuration, we
have determined the following differential resistances: At $T$=5.715 K, $%
dV/dI\simeq $0.1$\Omega $ for N=1 and $dV/dI\simeq $0.2 $\Omega $ for N=2,
which is in good agreement with the values given in Fig.~\ref{fig:dVdI}. At $%
T$=5.808 K, $dV/dI\simeq $0.12 $\Omega $ for N=1 and $dV/dI\simeq $0.24 $%
\Omega $ for N=2. In this case, the values shown in Fig.~\ref{fig:dVdI} are
higher although their ratio still remains approximately correct. These
calculations also point in the direction that the dissipative foot can be
interpreted in terms of vortex motion.

At higher currents, the $V(I)$ characteristics do not show a continuous
increase of voltage with increasing current but instead they are composed of
several voltage steps (see Fig.~\ref{fig:IV}). The transition to the normal
state of a 1D wire is produced by the generation of steady PSC's which
increase in number as the transport current grows. This dissipation
phenomenon correlates the differential resistance with the number of PSC's
created in the wire and therefore results in a $V(I)$ curve with
characteristic steps. Although the steps observed in the $V(I)$ curve in
Fig. \ref{fig:IV} are not steep like the ones reported for single 1D wires
\cite{Skocpol} and infinite wire networks \cite{Giroud92} interpreted by the
SBT theory, we believe that they are reminiscences of the resistive
transition of the quasi-one dimensional wires constituting the antidot
cluster. A non-strictly 1D character of the wires forming the antidot
cluster may be one of the reasons for the smoothness of the steps.


\section{Conclusions}

\label{sec:conc}

We have studied the transport properties of a Pb/Cu $2\times 2$ antidot
cluster (a microsquare with four antidots) by measuring the
superconducting/normal phase boundary, critical currents and $V(I)$
characteristics.

The Pb/Cu bilayer can be considered as a single superconducting entity with
superconducting parameters which are somewhat different than those of the
single Pb layer. The presence of the antidots leads to a characteristic
structure in the magnetoresistance and phase boundary caused by the
formation of well defined vortex configurations. By comparison of the
experimental results with calculations which approximate the cluster as a 1D
micro-network, we were able to identify the corresponding ground states.

The formation of particular vortex states, as the magnetic field is varied,
was also observed in the critical current versus field curves. Comparison
with a static model showed that the current injection lifts the degeneracy
and that the vortex states at larger current differ from the ground states,
observed in the phase boundary $T_c(B)$.

The dissipative processes were probed for the different vortex occupations
by means of $V(I)$ measurements. We find clear evidence for a stationairy
vortex motion at the onset of dissipation and the creation of phase slip
centers at larger voltages and currents.

\acknowledgments
We thank M. Kuprianov, A. L\'opez, H. Fink, A. Buzdin, M. Baert, J.G.
Rodrigo and V. Bruyndoncx for helpful discussions. The work has been
supported by the Belgian Inter-University Attraction Poles (IUAP), the
Flemish Concerted Action (GOA), VIS, and the National Fund for Scientific
Research (FWO-Vlaanderen) programs. One of us (T.P.) thanks the European
Training and Mobility of Researchers Program for financial support.

\begin{figure}[tbp]
\caption{Atomic force microscopy picture of ($a$) the $2\times 2$ antidot
cluster and ($b$) the reference sample without antidots.}
\label{fig:afm}
\end{figure}

\begin{figure}[tbp]
\caption{Schematic representation of the spatial dependence of the
superconducting order parameter at the interface between the superconductor
(S) and the normal material (N) of the Pb/Cu bilayers studied in this paper.
The two characterisic parameters for proximity induced superconductivity,
i.e. the coherence length in the normal layer, $\xi _N$ and the
extrapolation lenght, $b$, are indicated.}
\label{fig:prox}
\end{figure}

\begin{figure}[tbp]
\caption{Experimental S/N phase boundary of the $2\times 2$ antidot cluster
determined with a criterium of 10\% of the normal state resistance. Two
typical magnetoresistance curves are shown in the inset. The ac measuring
current was $1\mu$A.}
\label{fig:psbexp1}
\end{figure}

\begin{figure}[tbp]
\caption{Experimental S/N phase boundary of the reference sample determined
with a criterium of 10\% of the normal state resistance. The inset shows
four magnetoresistance curves measured at T=5.503 K, 5.482 K, 5.462 K and
5.443 K. The ac measuring current was 1 $\mu$A. The second critical magnetic
fields for the 1D transition ($B_{c2}^{1D}$), the 2D transition, ($%
B_{C2}^{2D}$), and the transition of the microsquare ($B_{c2}^{DOT}$) are
shown.}
\label{fig:psbref}
\end{figure}

\begin{figure}[tbp]
\caption{Schematic representation of the $2\times 2$ antidot cluster with
average dimensions indicated. The network approximation used in the
one-dimensional model is drawn with a dashed line. The dots denote the nine
nodes considered in the model.}
\label{fig:cluster}
\end{figure}

\begin{figure}[tbp]
\caption{($a$)Theoretical phase boundary, $\Delta T (\Phi/\Phi_{\circ})$,
obtained in the London limit of the Ginzburg-Landau theory (solid line). The
total energy for the six vortex configurations is shown by a dot line. The
dashed line indicates the non-stable 'parallel vortex configuration'. The
schematic representation of the vortex states in each parabolic branch is
also sketched by means of the corresponding $n_k$ quantum numbers. ($b$)
First period of the experimental phase boundary shown in Fig.~\ref
{fig:psbexp1} after subtraction of the parabolic background ($I_{ac}=1 \mu $%
A). ($c$) idem as ($b$) but for $I_{ac}=3 \mu$A. }
\label{fig:psbth1}
\end{figure}

\begin{figure}[tbp]
\caption{Comparison of the theoritical phase boundaries obtained for the
London (dashed line), Josephson junction array (solid line) and de
Gennes-Alexander (dot line) approaches.}
\label{fig:psbth2}
\end{figure}

\begin{figure}[tbp]
\caption{($a$)Theoretical phase boundary for an ideal one-dimensional
antidot cluster. ($b$) Theoretical phase boundary for an antidot cluster
considering the effects coming from the magnetic field penetration in the
quasi-one-dimensional wires. ($c$) Theoretical phase boundary for an antidot
cluster taking into account the effects induced by areal disorder. ($d$)
Theoretical phase boundary where the contributions shown in ($b$) and ($c$)
are both considered.}
\label{fig:psbdis}
\end{figure}

\begin{figure}[tbp]
\caption{Critical current density of the $2\times 2$ antidot cluster as
function of the reduced flux, $\Phi/\Phi _{\circ}$, at four temperatures. $%
(a)$ T=5.899 K, $(b)$ T=5.808 K, $(c)$ T=5.715 K and $(d)$ T=5.656 K.}
\label{fig:JcB}
\end{figure}

\begin{figure}[tbp]
\caption{Critical current versus the reduced flux per cell for ($a$) the
current-phase relation given by Eq.~\ref{eq:curphas} and ($b$) for a
sinusoidal current-phase relation. The vortex states generated in the $%
2\times 2$ antidot cluster at high transport currents are shown. ($c$)
Experimental critical current versus reduced flux per cell at T=5.808 K with
the background subtracted.}
\label{fig:critcurtheo}
\end{figure}

\begin{figure}[tbp]
\caption{$V(I)$ characteristics of the $2\times 2$ antidot cluster measured
at T=5.715 K for zero applied magnetic field and a magnetic field of 3 G and
12 G. In all the curves similar characteristic steps can be distinguished.}
\label{fig:IV}
\end{figure}

\begin{figure}[tbp]
\caption{Enlargement of the lower part of the $V(I)$ characteristics for a
reduced flux $\Phi = \Phi _{\circ}$, $\Phi = 0.37\Phi _{\circ}$ and $\Phi =
0.5\Phi _{\circ}$ at T=5.808 K and T=5.715 K.}
\label{fig:IV2}
\end{figure}

\begin{figure}[tbp]
\caption{Temperature dependence of the critical current measured at $B=0$
showing the same behaviour as the depairing critical current of 1D wires.}
\label{fig:IcT}
\end{figure}

\begin{figure}[tbp]
\caption{Differential resistance determined at the onset of the dissipative
foot of the $V(I)$ characteristics as a function of the reduced flux for
T=5.808 K and T=5.715 K. The solid line is a guide for the eyes showing the
two levels expected from the model of vortex motion induced by phase slips.
The dot lines indicate the $\Phi/\Phi _{\circ}$ values at which a change of
vortex configuration is observed experimentally.}
\label{fig:dVdI}
\end{figure}


\begin{references}
\bibitem[*]{byline}  Currently at Institut de Ciencia de Materials de
Barcelona (CSIC), Campus UAB, 08193 Bellaterra, Spain.

\bibitem{Mooij88}  Proceedings of the Nato Workshop on ''{\it Coherence in
Superconducting Networks}'', edited by J.E. Mooij and G.B.J. Sch\"on,
Physica B {\bf 152}(1988); Proceedings of the ICTP workshop on ''{\it %
Josephson Junction Arrays}, edited by H.A. Cerdeira and S.R. Shenoy, Physica
B {\bf 222} (1996).

\bibitem{Pannetier91}  B. Pannetier in {\it Quantum Coherence in Mesoscopic
Systems}, edited by B. Kramer, Plenum Press, New York, 1991.

\bibitem{Giroud92}  M. Giroud, O. Buisson, Y.Y. Wang, B. Pannetier and D.
Mailly, J. Low Temp. Phys. {\bf 87}, 683 (1992).

\bibitem{Vanderzant90}  H.S.J. van der Zant, M.N. Webster, J. Romijn and
J.E. Mooij, Phys. Rev B {\bf 42}, 2647 (1994).

\bibitem{Vanderzant94}  H.S.J. van der Zant, M.N. Webster, J. Romijn and
J.E. Mooij, Phys. Rev B {\bf 50}, 340 (1994).

\bibitem{Baert95}  M. Baert et al., Phys. Rev. Lett. {\bf 74}, 3269 (1995).

\bibitem{Bezryadin95}  M. Bezryadin and B. Pannetier, J. Low. Temp. Phys.
{\bf 98}, 251 (1995).

\bibitem{Rosseel96a}  E. Rosseel et al., Czech. J. Phys. {\bf 46} S 2, 885
(1996).

\bibitem{Moshchalkov96}  V.V. Moshchalkov et al., Phys. Rev. B {\bf 54},
7385 (1996).

\bibitem{Rosseel96b}  E. Rosseel et al., Phys. Rev. B {\bf 53}, R2983 (1996).

\bibitem{Puig96}  T. Puig, E. Rosseel, M. Baert, M.J. Van Bael, V.V.
Moshchalkov and Y. Bruynseraede, Appl. Phys. Lett.{\bf 70},3155 (1997).

\bibitem{Neerinck90}  D. Neerinck, K. Temst, H. Vanderstraeten, C. Van
Haesendonck, Y. Bruynseraede, A. Gilabert, and I. K. Schuller, Mat. Res.
Soc. Symp. Proc. Vol. {\bf 160} (1990).

\bibitem{Gilabert77}  A. Gilabert, Ann. Phys. t. 2, {\bf 203} (1977).

\bibitem{Meaden65}  G. T. Meaden, {\it Electrical Resistance of metals},
Plenum, New York, p. 4 (1965)

\bibitem{Biagi85}  K. R. Biagi, V. G. Kogan, J. R. Clem, Phys. Rev. {\bf 32}%
, 7165 (1985).

\bibitem{DeGennesbook}  P. G. de Gennes, {\it Superconductivity of metals
and alloys}, Addison-Wesley Publishers (1966).

\bibitem{DeGennes64}  P. G. de Gennes, Rev. Mod. Phys. {\bf 36}, 225 (1964).

\bibitem{Neerinck91}  D. Neerinck, Ph. D. thesis, Katholieke Universiteit
Leuven, Leuven (1991).

\bibitem{Tinkham75}  M. Tinkham, {\it Introduction to Superconductivity},
Mc. Graw-Hill, New-York (1975).

\bibitem{Moshch95}  V.V. Moshchalkov, L. Gielen, C. Strunk, R. Jonckheere,
X. Qiu, C. Van Haesendonck, and Y. Bruynseraede, Nature {\bf 373}, 319
(1995).

\bibitem{Chi94}  C.C. Chi, P. Santhanam, S.J. Wind, M.J. Brady, and J.J.
Bucchignano, Phys. Rev. B {\bf 50}, 3487 (1994).

\bibitem{Chi92}  C.C. Chi and P. Santhanam, J. Low Temp. Phys. {\bf 88}, 163
(1992).

\bibitem{VVM}  V.V.Moshchalkov et al., Physica Scripta {\bf T55},168 (1994)

\bibitem{DeGennesalexander}  P.G. de Gennes, C.R. Acad. Sci. B {\bf 292},
279 (1981); S. Alexander, Phys. Rev. B {\bf 27}, 1541 (1983).

\bibitem{Fink91a}  H.J. Fink and S.B. Haley, Phys. Rev. Lett. {\bf 66}, 216
(1991).

\bibitem{Fink91b}  H.J. Fink, O. Buisson and B. Pannetier, Phys. Rev. B {\bf %
43}, 10144 (1991).

\bibitem{Itzler90}  M.A. Itzler et al., Phys. Rev. B {\bf 42}, 8319 (1990).

\bibitem{Benz88}  S.P. Benz, M.G. Forrester, M. Tinkham and C.J. Lobb, Phys.
Rev. B {\bf 38}, R2869 (1988).

\bibitem{Harlingen93}  S.V. Rao and D.J. Van Harlingen, Phys. Rev. B {\bf 48}%
, 12853 (1993).

\bibitem{Likharev79}  K.K. Likharev, Rev. Mod. Phys. {\bf 51}, 101 (1979).

\bibitem{Press90}  W.H. Press et al. in {\it Numerical Recipes in Pascal},
Cambridge University Press, Cambridge, 1990.

\bibitem{Chung89}  J.S. Chung, K.H. Lee and D. Stroud, Phys. Rev. B {\bf 40}%
, 6570 (1989).

\bibitem{Skocpol}  W.J. Skocpol, M.R. Beasley and M. Tinkham, J. Low Temp.
Phys. {\bf 16}, 145 (1974).

\bibitem{Kopnin84}  B.I. Ivlev and N.B. Kopnin, Adv. Phys. {\bf 33}, 47
(1984).

\bibitem{Langer}  J.S. Langer and V. Ambegoakar, Phys. Rev. {\bf 164}, 498
(1967).

\bibitem{Phillips94}  J.R. Phillips, H.S.J. van der Zant and T.P. Orlando,
Phys. Rev. B {\bf 50}, 9380 (1994).

\bibitem{Bardeen}  J. Bardeen and M.J. Stephen, Phys. Rev. {\bf 140}, A1197
(1965).

\bibitem{Orlando91}  T.P. Orlando, J.E. Mooij and H.S.J. van der Zant, Phys.
Rev. B {\bf 43}, 10218 (1991).

\bibitem{Benoist96}  R. Benoist and W. Zwerger, Z. Phys.B {\bf 103}, 377
(1997)
\end{references}
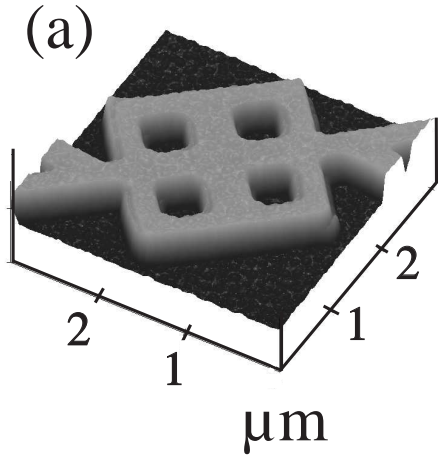
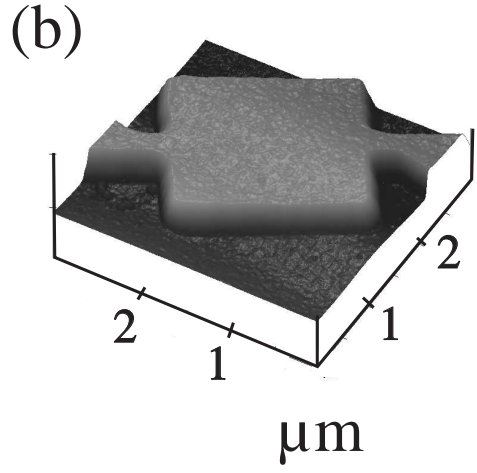
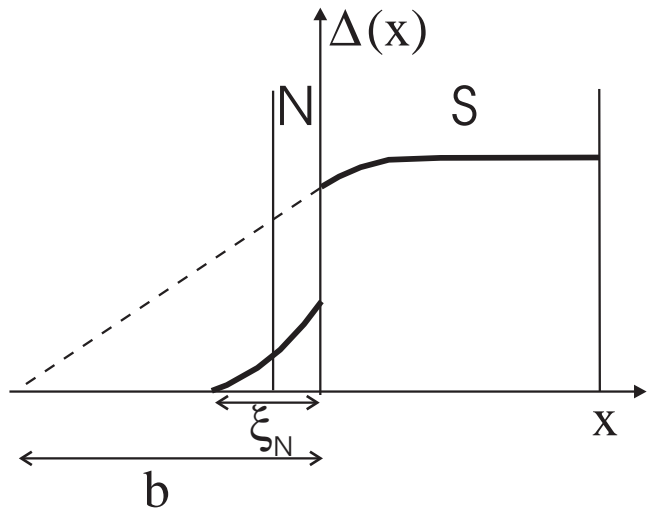
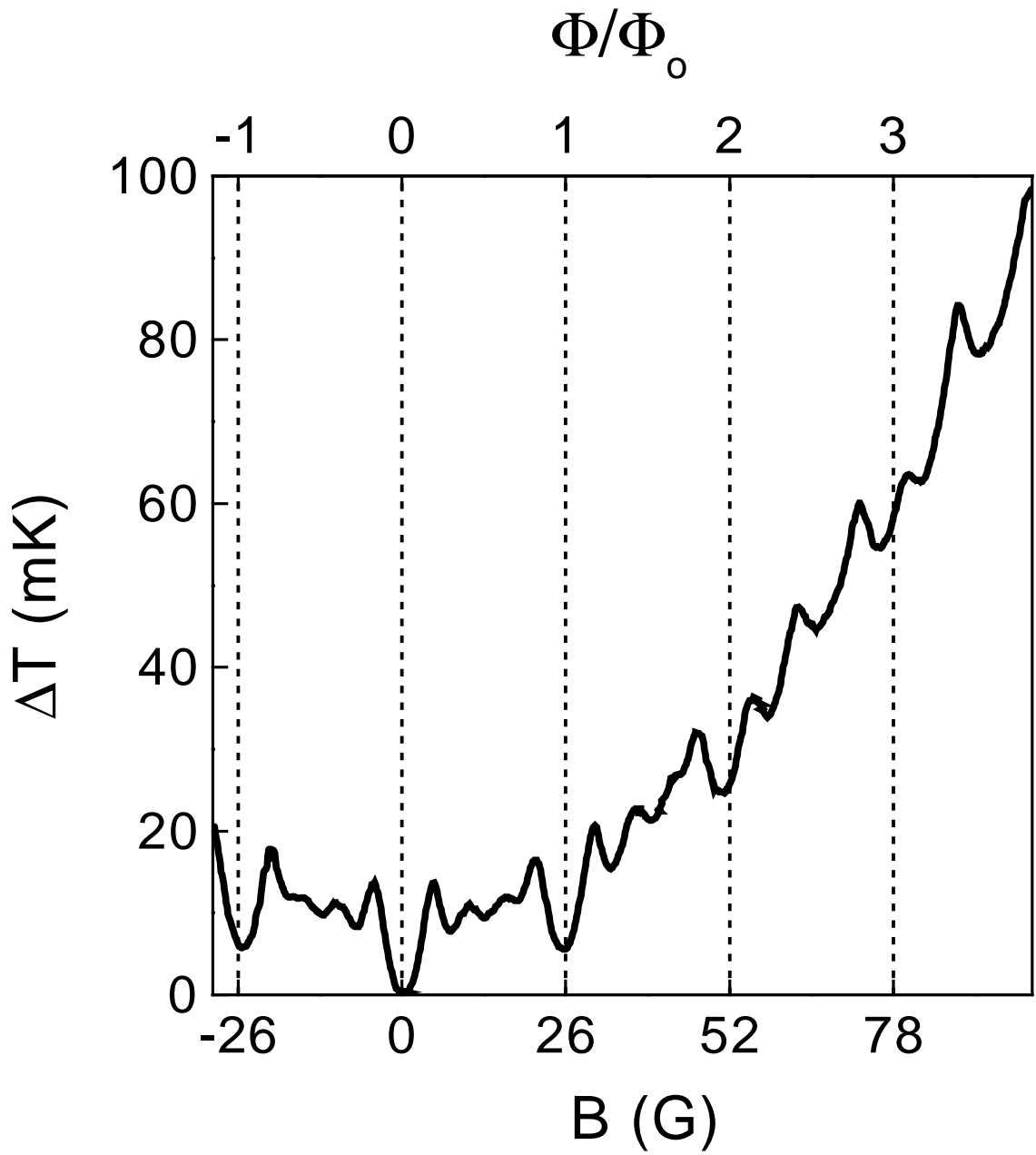
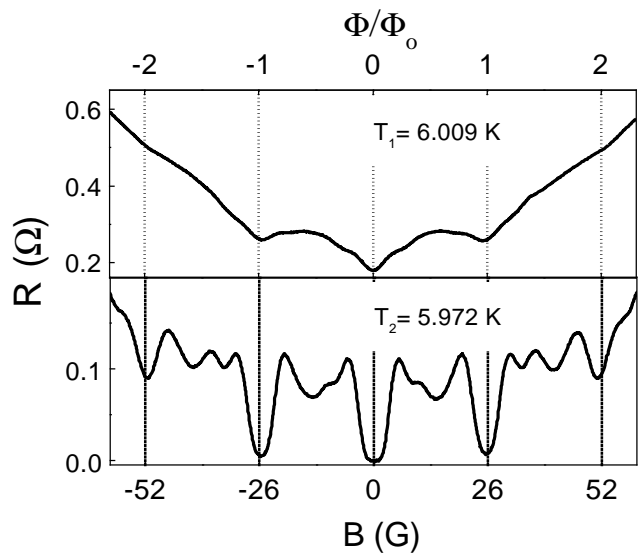
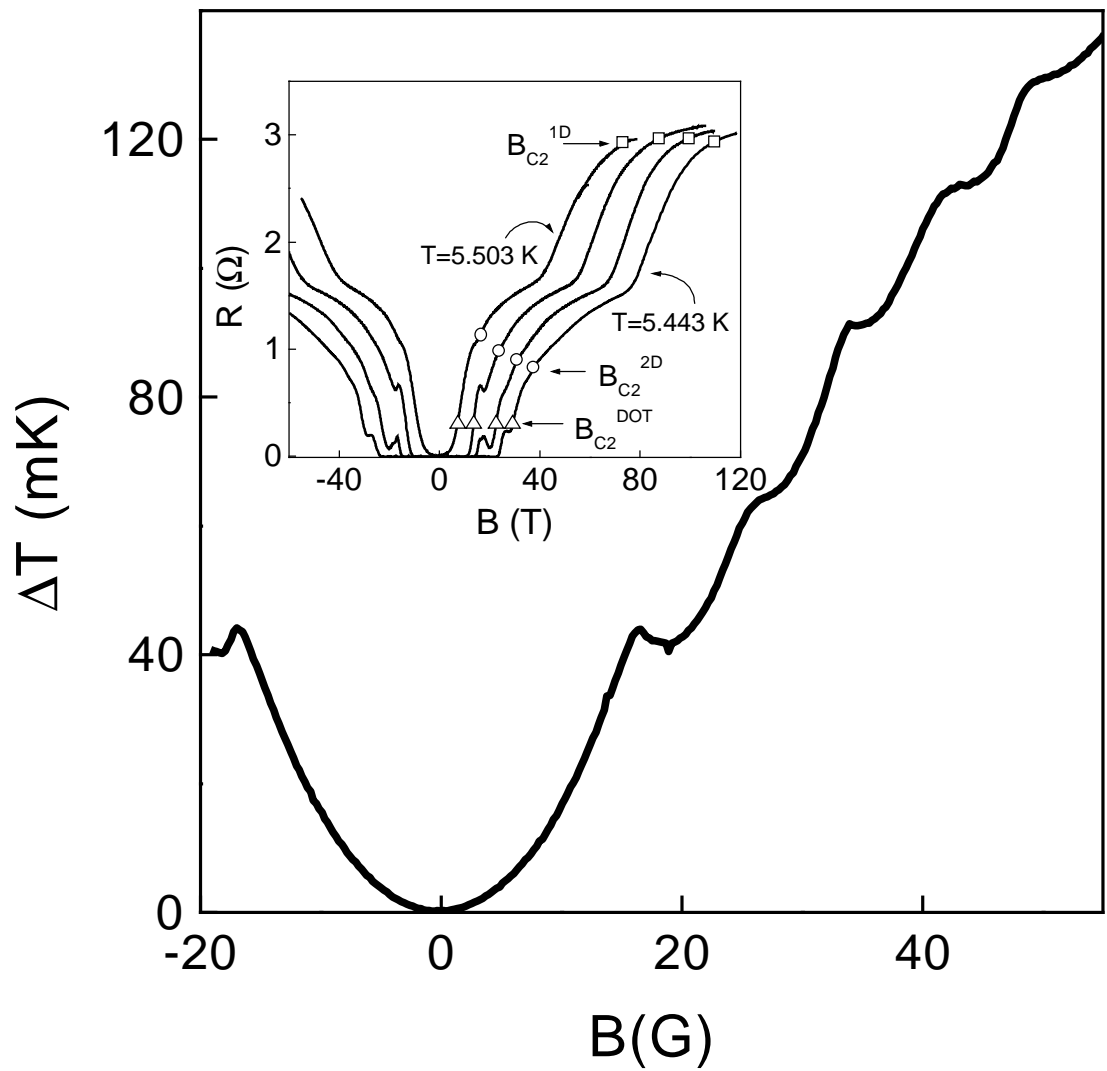
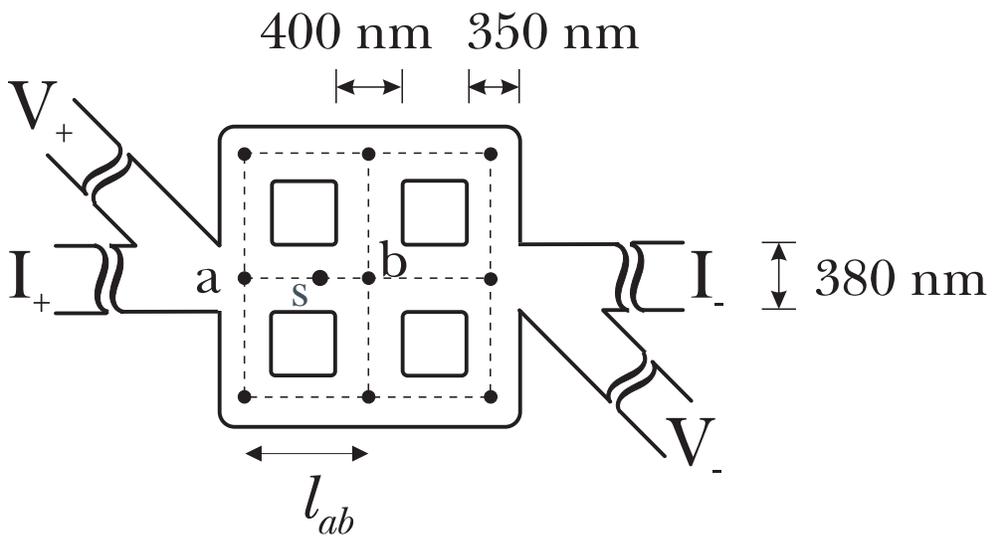
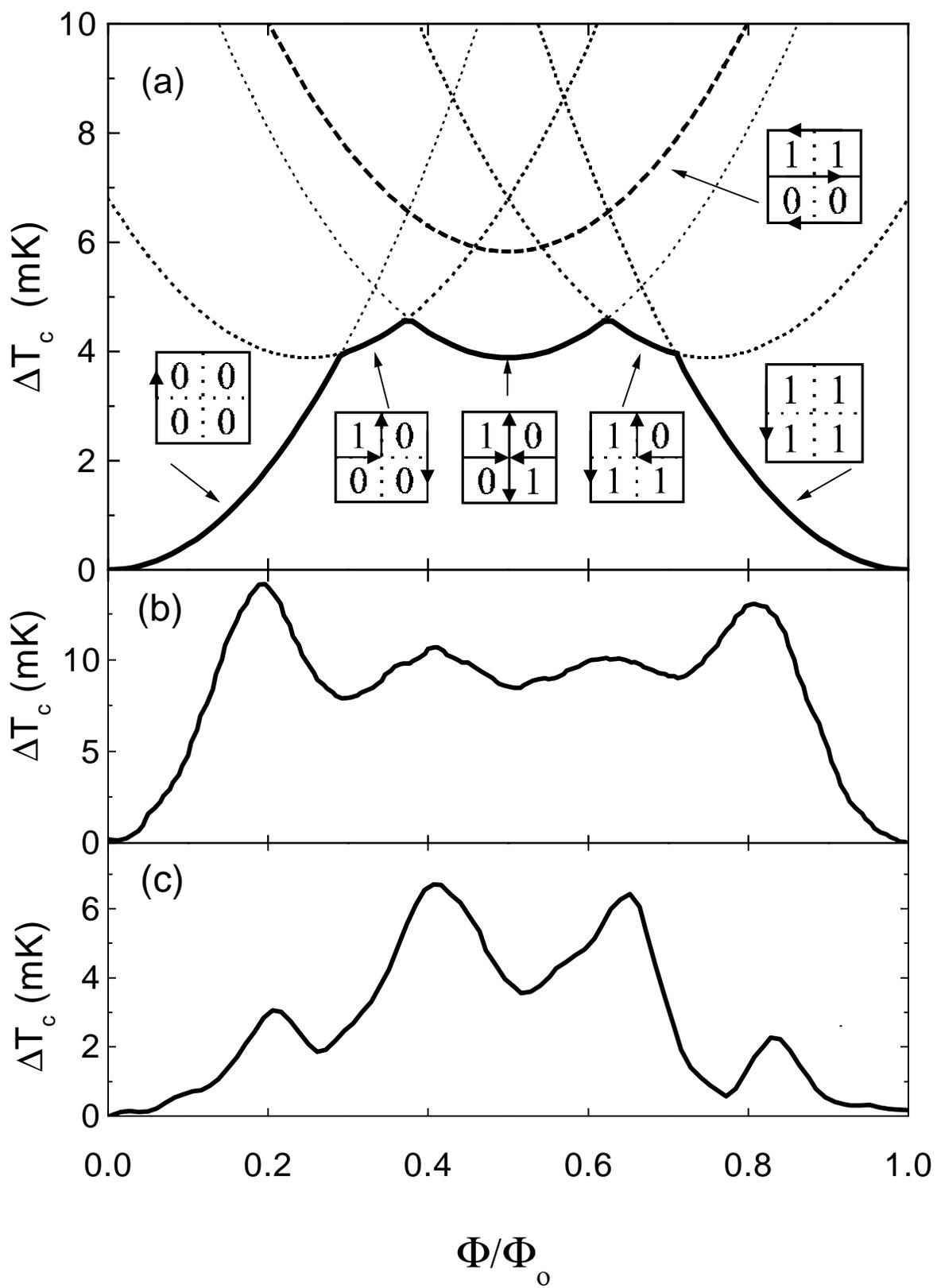
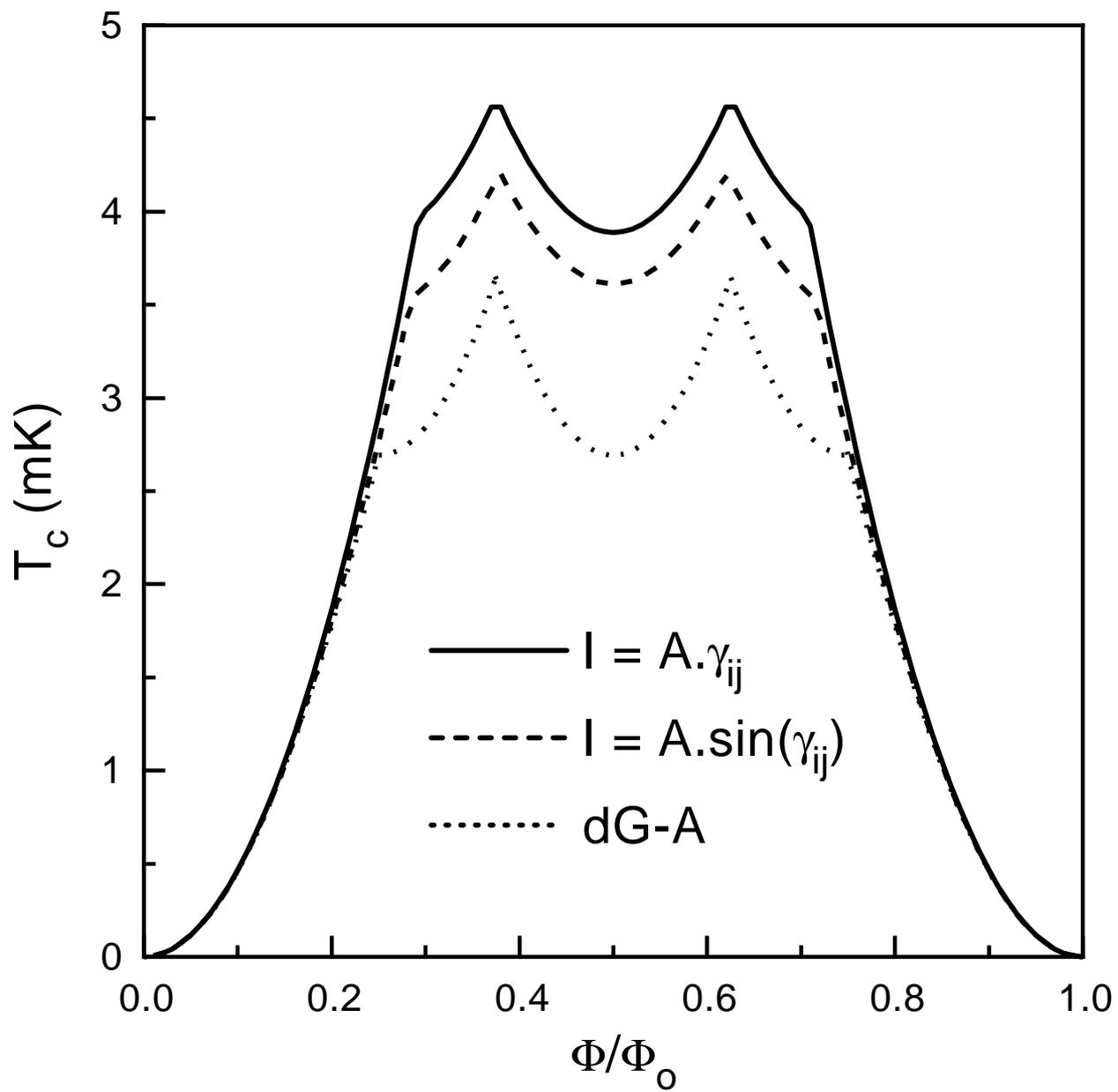
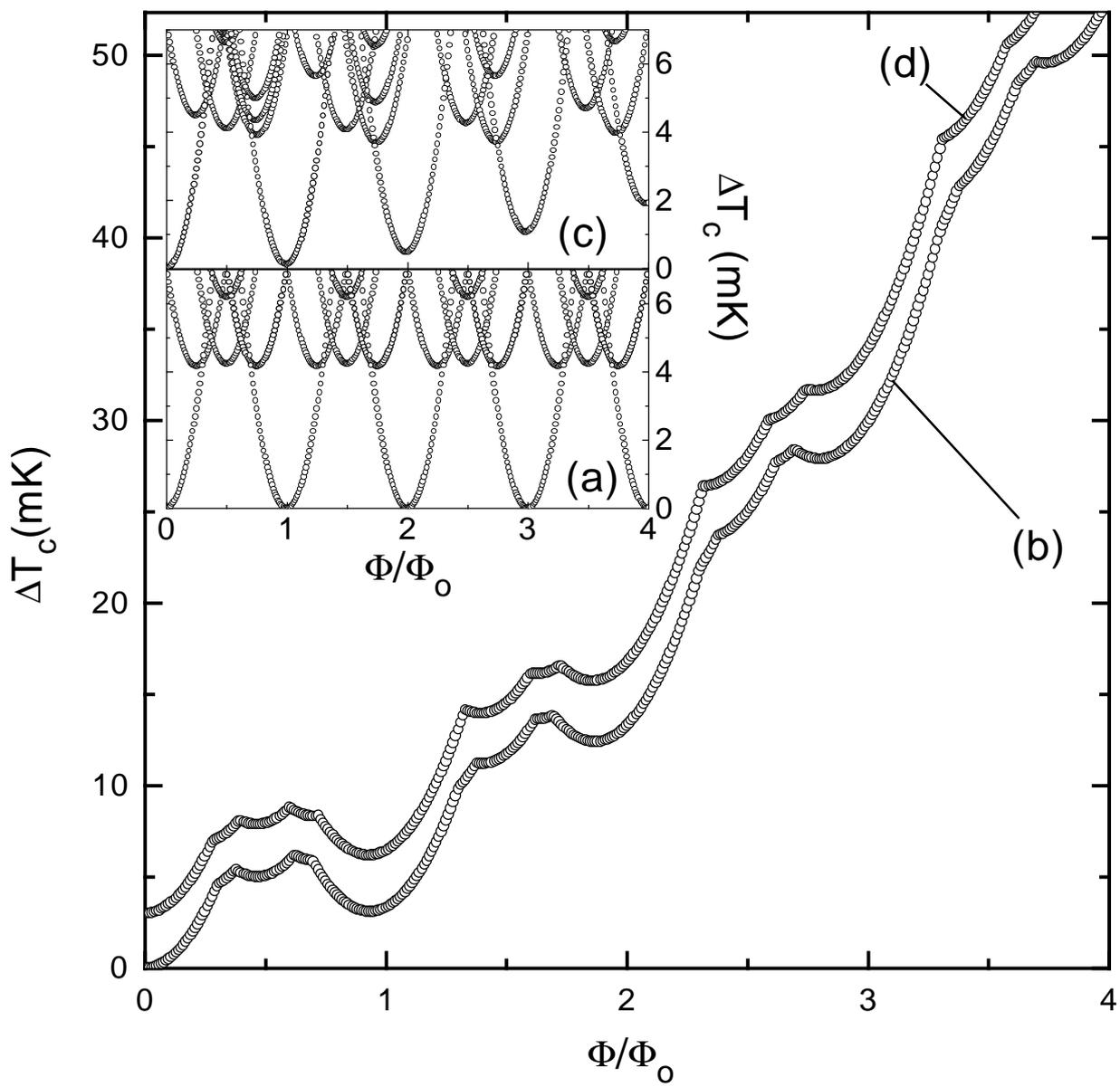
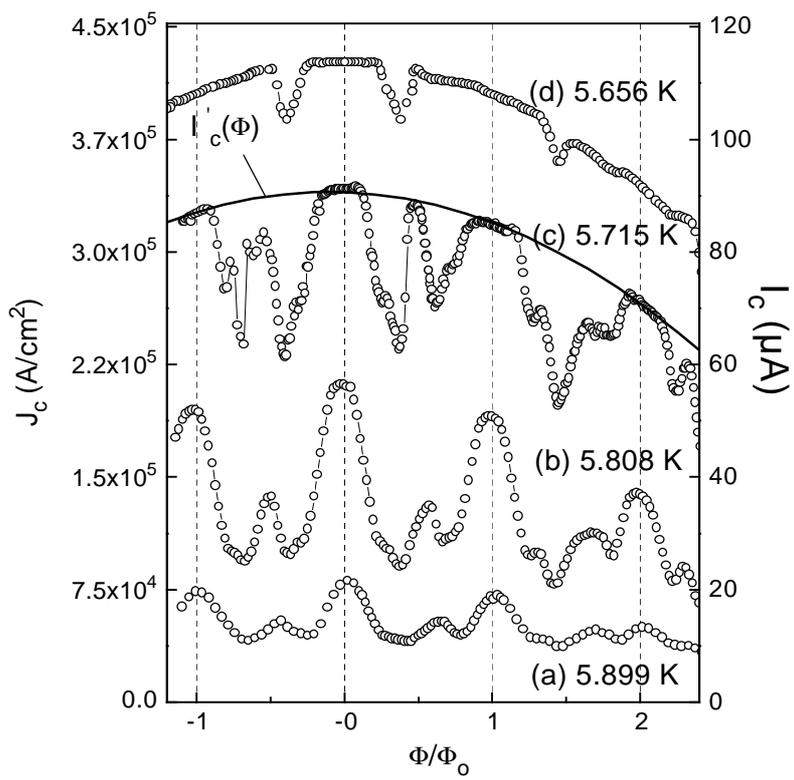
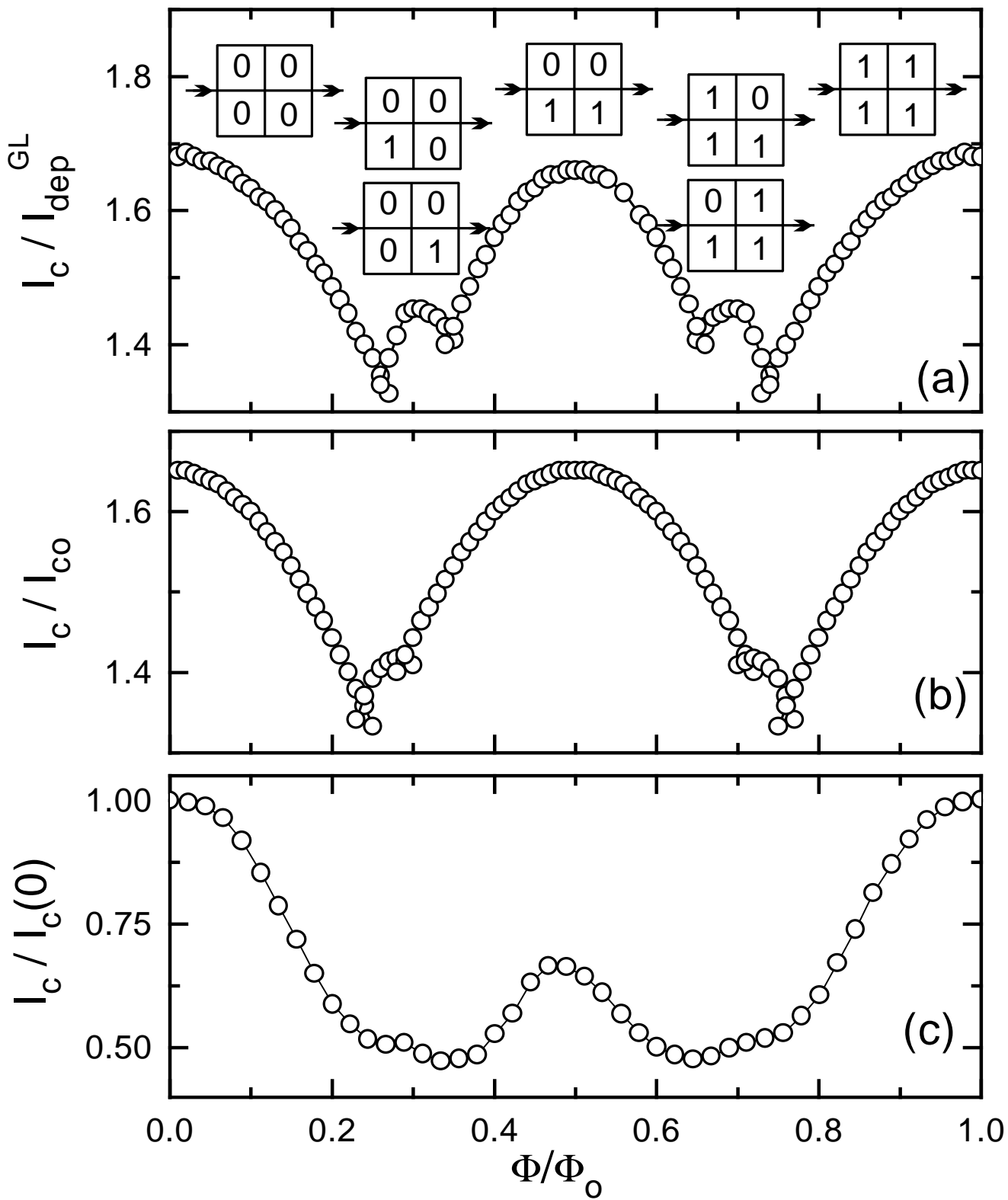
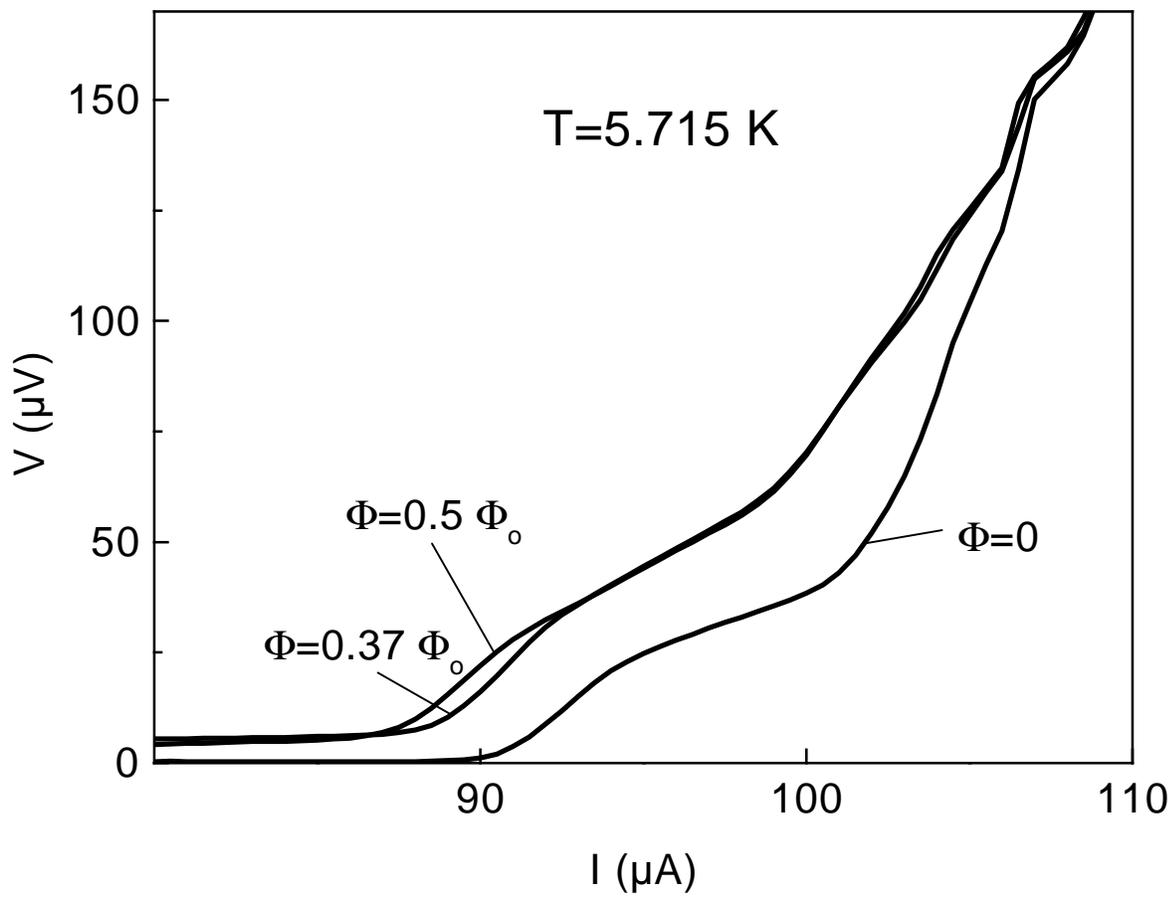
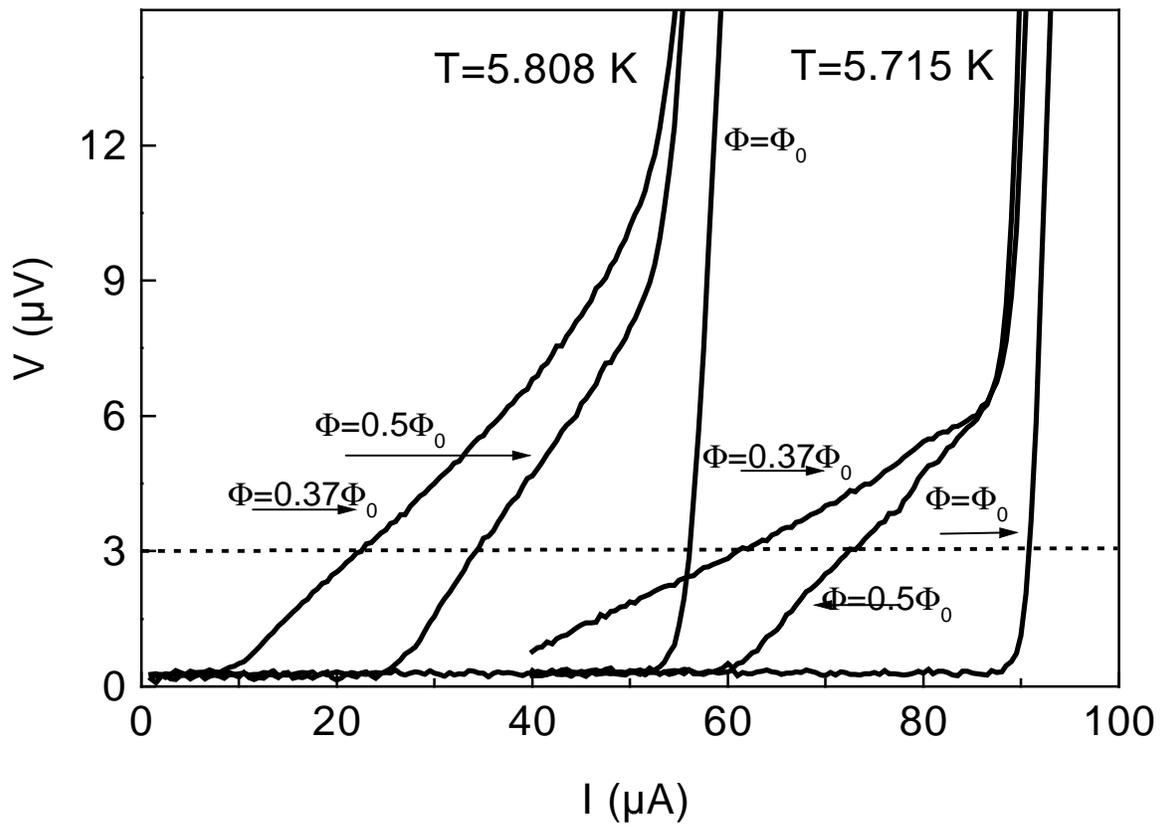
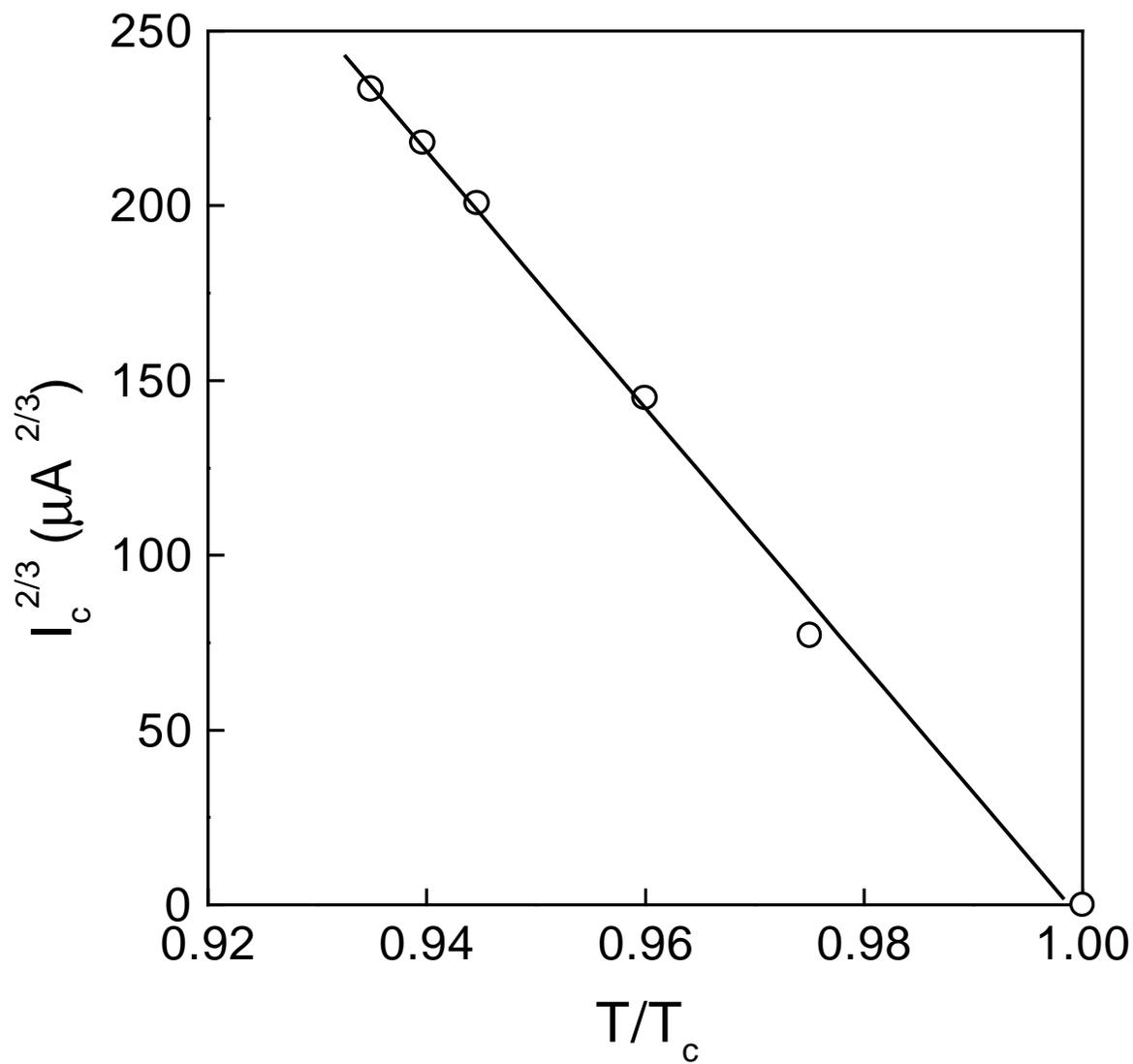
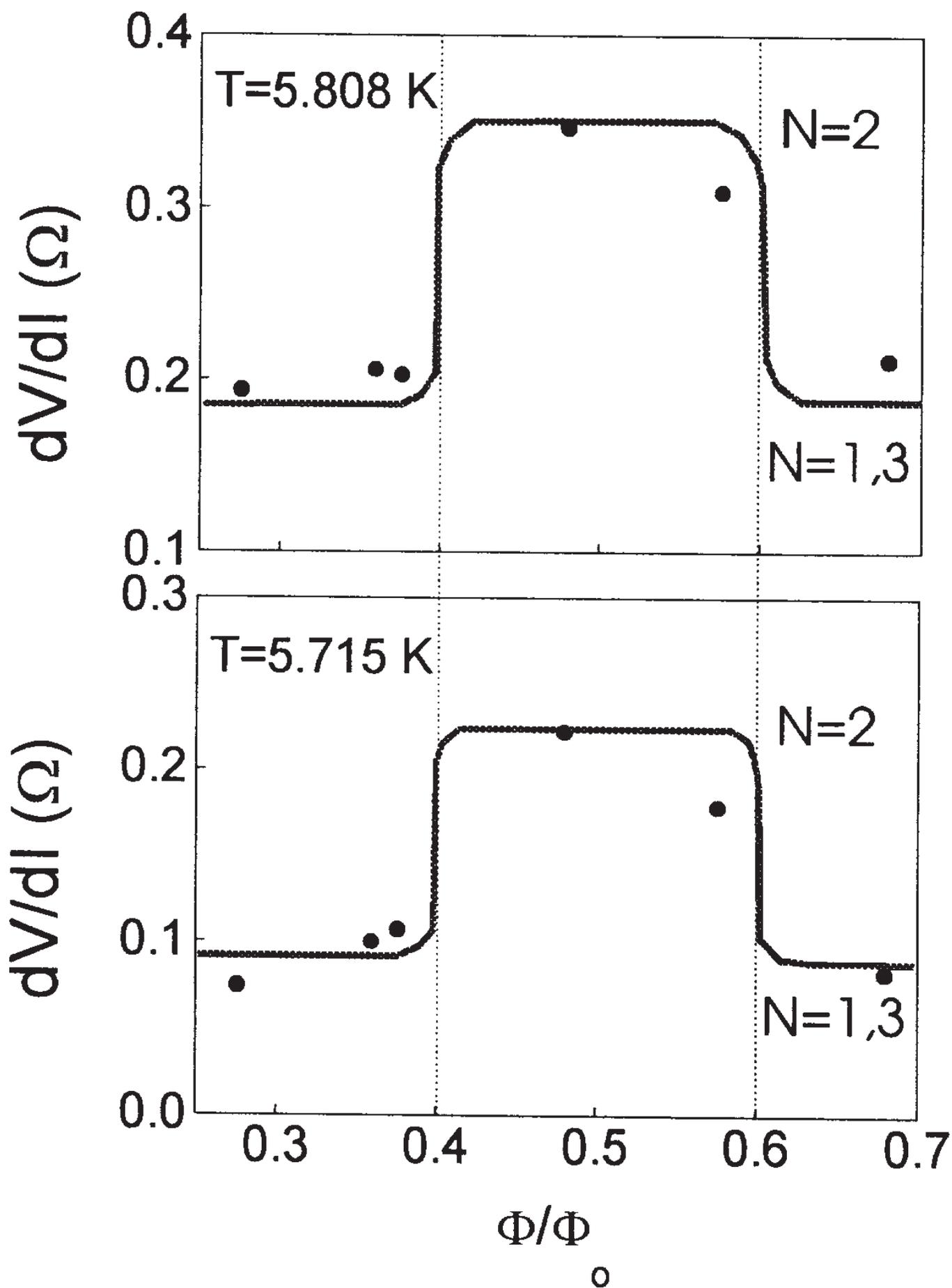
\end{document}